\documentclass[preprint,12pt]{elsarticle}

\usepackage{amssymb}
\usepackage{graphicx}
\usepackage{dcolumn}
\usepackage{bm}
\usepackage{url}
\usepackage[utf8]{inputenc}
\usepackage[T1]{fontenc}
\usepackage[draft]{todonotes}
\usepackage{threeparttable}
\usepackage[abbreviations]{glossaries-extra}
\usepackage{enumerate}
\usepackage{mathptmx}
\usepackage{etoolbox}
\usepackage{color}
\usepackage{CJKutf8}
\usepackage{xcolor}
\usepackage{soul}

\soulregister\cite7
\soulregister\ref7
\soulregister\eqref7
\soulregister\pageref7 

\newabbreviation{lbm}{LBM}{lattice Boltzmann method}
\newabbreviation{lb}{LB}{lattice Boltzmann}
\newabbreviation{ns}{NS}{Navier-Stokes}
\newabbreviation{pde}{PDE}{partial differential equation}

\journal{International Journal of Multiphase Flow}
\begin{document}

\begin{frontmatter}



\title{Thermal effects connected to crystallization dynamics: a lattice Boltzmann study}


\author[inst1]{Q. Tan}

\author[inst1,inst3]{S.A.~Hosseini}

\author[inst2]{A.~Seidel-Morgenstern}
            
\author[inst1]{D.~Th\'evenin}

\author[inst2]{H.~Lorenz}

\affiliation[inst1]{organization={Laboratory of Fluid Dynamics and Technical Flows, University of Magdeburg ``Otto von Guericke''},
            city={Magdeburg},
            postcode={39106}, 
            country={Germany}}

\affiliation[inst2]{organization={Max Planck Institute for Dynamics of Complex Technical Systems (MPI DCTS)},
            city={Magdeburg},
            postcode={39106}, 
            country={Germany}}
            
\affiliation[inst3]{organization={Department of Mechanical and Process Engineering},
            city={ETH Z\"urich},
            postcode={8092}, 
            country={Switzerland}}

\begin{abstract}
The possible impact of temperature differences during crystal growth is investigated in this study. The organic molecule considered is mandelic acid, an important component for the pharmaceutical industry. The productivity of generating pure mandelic acid crystals are largely determined by the growth process. Reaction conditions, purity of the components, supersaturation, temperature, but possibly also temperature gradients play a central role during crystal growth. In this study a numerical model based on a hybrid solver combining the lattice Boltzmann method with finite differences is developed to model the crystallization dynamics of (S)-mandelic acid (S-ma) taking quantitatively into account temperature effects. At first, the fourth-order finite-difference method used to model energy and species conservation is validated. Then, comparisons are carried out regarding temperature changes within the single-crystal growth cell. In practice, the molar heat generation at the crystal interface shows only a small effect on the temperature field in the surrounding domain, with temperature differences below $1.5$ degree. Finally, the study is extended to investigate the impact of forced convection on the crystal habits while taking into account temperature differences.
\end{abstract}


\end{frontmatter}

\section{Introduction}
Mandelic acid and its derivatives are frequently used compounds in the pharmaceutical industry. It exists as two pure enantiomers and in the racemic form, with strong consequences on its pharmacological properties~\cite{chlebus2006mandelic}. It is also used for the further organic synthesis of pharmaceuticals, such as esters of mandelic acid generating homatropine for eye drops. Mandelic acid is well known for its anti-aging effects on the skin along with antibacterial functions in treating acne~\cite{emel2018experimental}. Furthermore, the manufacture of many rubbers, adhesives, and plastic materials requires mandelic acid as an intermediate substance.
(S)- and (R)-mandelic acid are the two enantiomeric forms. Enantiopure substances are required for most pharmaceutical applications~\cite{brittain2002mandelic}. Crystallization is widely used for the separation of enantiomers relying on classical resolution, or preferential crystallization approaches~\cite{lorenz2014processes}. During crystallization, essential properties of the crystalline products (e.g., purity, shape, sizes~\cite{briesen2006simulation}) are determined by the growth process, which again depends on the conditions within the crystallizer. The reaction conditions, such as supersaturation, temperature, other components possibly present in the solution (impurities, additives) play a central role for crystal growth. Many experimental studies have been conducted concerning crystallization-based enantio-separation processes. Of particular interest for the present work are measurements regarding growth kinetics of mandelic acid, e.g.~\cite{alvarez2004online,lorenz2014processes, coquerel2006preferential, gansch2021continuous, gou2012investigation, perlberg2005crystal, srisanga2015crystal,codan2013growth}. Most of the studies carried out up to now relied on the assumption of a perfectly homogeneous temperature during crystallization. Typically, experimental temperature measurements rely on a single sensor (point measurement), so that possible temperature gradients could not be tracked. Since only small temperature differences are expected, experimental investigations regarding temperature effects during crystal growth would be challenging and costly.\\
Numerical simulations using accurate and efficient algorithms can in this case complement or replace such experiments and provide corresponding answers. In recent years, much effort has been put on developing mathematical models and numerical algorithms suitable for describing crystal habit and size of crystals~\cite{karma1998quantitative,younsi2016anisotropy}, also for enantiopure (S)-mandelic acid~\cite{tan2021modeling,tan2022mandelic}. The phase-field method has become one of the most popular approaches to simulate crystal growth. It is a powerful tool for modeling structural evolution of materials and crystals~\cite{vakili2020multi,subhedar2020diffuse, schiedung2020simulation}. It is now widely used to investigate solidification~\cite{boettinger2002phase,nestler2002phase} and grain growth~\cite{chen1994computer,takaki2016two,karma1998quantitative,tourret2017grain}. The phase-field approach has also been used in combination with the lattice Boltzmann method, now widely recognized as an efficient alternative to classical tools, to simulate solidification processes~\cite{younsi2016anisotropy,lin2014three,wang2019brief,rojas2015phase,m2019non}. This approach can reproduce numerically the solid-liquid interface interactions and the hydrodynamic effects affecting the habits of growing crystals~\cite{Henniges2017,medvedev2006influence,sakane2018three,chakraborty2007enthalpy,tan2021modeling,tan2022mandelic}. While widely used in the literature for hydrodynamic simulations the lattice Boltzmann method is known to suffer from Gibbs-type oscillations near sharp interfaces and instability issues in the limit of vanishing diffusion coefficients. Furthermore, the classical passive-scalar lattice Boltzmann solvers can not take into account flows with variable density and/or specific heat capacity. For such flows the models need to be extended, see for instance \cite{hosseini2019lattice}. In such cases an interesting alternative is to replace the solvers for the scalar fields, e.g. species and temperature, with classical finite-difference solvers with discontinuity-capturing schemes for the advection term. The corresponding finite-difference solvers are then coupled to a lattice Boltzmann approach for describing hydrodynamics. Such hybrid approaches have been increasingly used in the past years for applications such as combustion, see for instance~\cite{hosseini2019hybrid,hosseini2020low,hosseini2022low}.\\
In the present work the crystal growth of (S)-mandelic acid is studied in detail using a hybrid lattice Boltzmann/finite-difference method under different reaction conditions and taking into account temperature difference changes; additionally, possible convection (sometimes also called ventilation) effects will be considered. At the difference of previous works, the enthalpy production due to mandelic acid lattice integration is included in the model and an energy balance equation is solved in the whole domain; in this manner, the effects of possible temperature gradients within the crystallizer are fully taken into account. In companion experiments, well-characterized seed crystals must be produced from supersaturated aqueous (S)-mandelic acid solutions. The single grain growth is then tracked, the growing crystal being inserted into a dedicated measurement cell. For the present studies focusing on thermal effects: (1) the growth rate of the crystal was investigated for different crystallization temperatures in the growth cell; (2) numerically, heat generation is taken into account at the crystal interface and temperature changes are solved for within the entire growth cell. Finally, (3) the impact of convection effects on crystal habit is studied at different Reynolds numbers. Baffles are additionally placed in the cell to support symmetrical crystal growth.

\section{\label{sec:level2}Numerical methods}

\subsection{Governing equations}

\subsubsection{\label{sec:level2.1}Diffuse-interface formulation: governing equations}

In the phase-field method solid growth dynamics are expressed via a non-dimensional order parameter, $\phi$, going from (+1) in the solid to (-1) in the pure liquid phase. The space/time evolution equations are written as~\cite{jeong2001phase,beckermann1999modeling}:
\begin{multline}
    \tau_0 a_s^2(\textbf{n}) \frac{\partial \phi}{\partial t} = W_0^2  \bm{\nabla} \cdot \left(a_s^2(\textbf{n})\right) \bm{\nabla} \phi +  W_0^2 \bm{\nabla} \cdot \left (|\bm{\nabla} \phi|^2  \frac{\partial[a(\textbf{n})^2]}{\partial \bm{\nabla} \phi}\right )\\
    + \frac{(\phi - \phi^3) + (\lambda_1 U + \lambda_2 \theta)  (1 - \phi^2)^2}{\tau_0} ,
  \label{a}
\end{multline}
and regarding normalized supersaturation $U$:
\begin{equation}
    \frac{\partial U}{\partial t} + \left( \frac{1-\phi}{2}\right) \bm{u} \cdot \bm{\nabla} U =D \bm{\nabla} \cdot \left(q(\phi) \bm{\nabla} U \right) - \frac{1}{2} \frac{\partial \phi}{\partial t},
     \label{ccc}
\end{equation}
and temperature $\theta$:
\begin{equation}
    \frac{\partial T}{\partial t} + \left( \frac{1-\phi}{2}\right) \bm{u} \cdot \bm{\nabla} T  =\frac{1}{\rho \widetilde{c_p}}\bm{\nabla}\rho \widetilde{c}_p\widetilde{\kappa}\cdot\bm{\nabla} T
    + \frac{1}{2} \frac{\Delta H_{\rm cryst}}{\widetilde{c}_p}\frac{\partial \phi}{\partial t},
     \label{b}
\end{equation}

where $\tau = \tau_0 a_s^2(\textbf{n})$. The coefficient $\lambda_1$ and $\lambda_2$ describes the strength of the coupling between the phase-field and the supersaturation field $U$,the temperature field $T$, respectively. $\theta = (T - T_1)/T_1$ is the normalized temperature in the phase field equation. $T_1$ is the constant temperature in the growth cell(see Fig.~\ref{ba}). Both $U$ and $\theta$ contribute to the driving force for the crystal growth. The parameter $\lambda_1 = \frac{\tau_0}{W_0^2}\cdot \frac{D}{a_2}$, where $D$ is the diffusion coefficient of the solution and $a_2=0.6267$~\cite{ramirez2004phase}. Here $\lambda_1=3.0$~\cite{tan2022mandelic}. The parameter $\tau_0$ denotes the characteristic time and $W_0$ the characteristic width of the diffuse interfaces. In Eq.~(\ref{a}), the quantity $\textbf{n} = - \frac{\bm{\nabla} \phi}{\left| \bm{\nabla} \phi \right|}$ is the unit vector normal to the crystal interface pointing from solid to fluid, while $a_s(\textbf{n})$ is the surface tension anisotropy function. In the context of the hexagonal mandelic acid crystal growth, this quantity is defined as~\cite{karma1998quantitative}:
\begin{equation}
    a_s(\textbf{n}) = 1 + \epsilon_s \cos(6 \varphi),
\end{equation}
\newabbreviation{rhs}{RHS}{right hand side}
\newabbreviation{lhs}{LHS}{left hand side}
with $\varphi = \arctan(n_y/n_x)$ considering the two spatial dimensions $x$ and $y$. The numerical parameter $\epsilon_s$ characterizes the anisotropy strength, and is set in the present study to $\epsilon_s = 0.05$ following~\cite{karma1996phase}. The term $(\phi - \phi^3)$ is the derivative of the double-well potential. The last term in Eq.~(\ref{a}) is a source term accounting for the coupling between supersaturation $U$, temperature $\theta$, and order parameter $\phi$. There, $(1 - \phi^2)^2$ is an interpolation function minimizing the bulk potential at $\phi = \pm 1$.\\
In Eq.~(\ref{ccc}), $\bm{u}$ denotes the local fluid velocity while $q(\phi) = (1 - \phi)$ is a function canceling out diffusion within the solid. As a consequence, solute transport is assumed to take place only within the fluid phase (one-sided model). The parameter $D$ is the diffusion coefficient of (S)-mandelic acid in water. Normalized supersaturation $U$ is later defined in Eq.~(\ref{U}); its transport equation is given by Eq.~(\ref{ccc}).\\
In Eq.~(\ref{b}), $\tilde{\kappa}$ is the thermal diffusivity in the single-crystal growth cell, which is defined as:
\begin{equation}
    \tilde{\kappa}  = \frac{(1-\phi)\kappa_L + (1+\phi)\kappa_S}{2},
\end{equation}
where $\kappa_L$ is the thermal diffusivity for the solution and $\kappa_s$ is for the solid. The quantity $\tilde{\kappa}$ tracks the different values of thermal diffusivity between the liquid and solid phases. Similarly, for specific heat capacity:
\begin{equation}
    \widetilde{c}_p = \frac{(1-\phi)c_{p,L} + (1+\phi)c_{p,S}}{2}.
\end{equation}
where $c_{p,L}$ is the specific capacity for the liquid and $c_{p,S}$ is for the crystal. The quantity $\Delta H_{cryst}$ represents the energy difference per mole of S-ma between the crystal solid and aqueous phases (see Table~\ref{phy}).

\subsubsection{Flow field formulation}
The mass conservation (or continuity) equation reads for this incompressible system:
\begin{equation}
    \bm{ \nabla }\cdot \left[ \frac{1-\phi}{2}\bm{u}\right] =0,
    \label{aaa}
\end{equation}
where $\bm{u}$ is the velocity of the flow field. The momentum conservation equation is as follows:
\begin{multline}
     \frac{\partial}{\partial t} \left[ \frac{1-\phi}{2} \bm{u}\right] + \bm{u} \cdot \bm{\nabla} \left[ \frac{1-\phi}{2} \bm{u} \right] + \left( \frac{1-\phi}{2} \right) \frac{\bm{\nabla} P}{\rho_0}\\
    = \nu \bm{\nabla^2} \left[ \frac{1-\phi}{2} \bm{u} \right] - \nu \frac{h(1+\phi)^2 (1-\phi)}{4W_0^2} \bm{u},
  \label{bbb}
\end{multline}
where $t$ is time, $P$ pressure, $\rho_0$ the liquid phase density, $\nu$ kinematic viscosity, $W_0$ interface thickness and $h$ a constant (equal to $2.757$) that ensures that the interface shear is correct for a simple shear flow~\cite{beckermann1999modeling}.

\subsection{\label{sec:level2.2}Numerical methods}

\subsubsection{Flow field solver with \gls{lbm}}
The flow field behavior (described by the incompressible Navier-Stokes and continuity equations) is modeled using the classical LB formulation consisting of the now-famous stream-collide operators:
\begin{equation}
    f_\alpha \left( \bm{x}+\bm{c}_\alpha \delta t, t+\delta t\right) - f_\alpha \left( \bm{x}, t\right) = \delta t \Omega_\alpha\left( \bm{x}, t\right) + \delta t\bm{F},
\end{equation}
where $\bm{F}$ is the external force. Here, $\bm{F}$ is used to represent the interaction with the solid phase following~\cite{beckermann1999modeling}:
\begin{equation}
    F = -\frac{h\eta_f (1+\phi)^2 (1-\phi)\bm{u}}{4W_0^2}
\end{equation}
where $h$ is a dimensionless constant, chosen as $h = 2.757$~\cite{beckermann1999modeling}. Due to the absence of fluid velocity within the solid crystal, the velocity variable $\bm{u}$ is updated as:
\begin{equation}
    \bm{u^*} = \frac{(1-\phi)}{2}\bm{u},
\end{equation}
and the corrected fluid velocity $\bm{u^*}$ is used in the equilibrium distribution function~\cite{beckermann1999modeling}. The collision operator $\Omega_\alpha$ follows the linear Bhatnagar-Gross-Krook (BGK) approximation:
\begin{equation}
    \Omega_\alpha = \frac{1}{\tau}\left[f^{(eq)}_\alpha  - f_\alpha\right],
\end{equation}
\newabbreviation{edf}{EDF}{equilibrium distribution function}
where $f_\alpha^{(eq)}$ is the discrete isothermal \gls{edf} defined as:
\begin{equation}
    f_\alpha^{(eq)} = \rho w_\alpha\sum_i \frac{1}{i! c_s^{2i}} a^{(eq)}_i(\bm{u}):\mathcal{H}_{i}(\bm{c}_\alpha),
\end{equation}
where $a^{(eq)}_i$ and $\mathcal{H}_{i}(\bm{c}_\alpha)$ are the corresponding multivariate Hermite coefficients and polynomials of order $i$, with $c_s$ the lattice sound speed corresponding to the speed of sound at the stencil reference temperature, and $w_\alpha$ the weights associated to the Gauss-Hermite quadrature~\cite{shan2006kinetic}. Further information on the expansion along with detailed expressions of the \gls{edf} can be found in~\cite{shan2006kinetic,hosseini2019extensive,hosseini2020development}. In the present work, an extended range of stability is obtained by using a central Hermite multiple relaxation time (MRT) implementation; corresponding details can be found in~\cite{hosseini2021central}. The relaxation time $\tau$ is tied to the fluid kinematic viscosity as:
\begin{equation}
    \tau = \frac{\nu}{c_s^2} + \frac{\delta t}{2}.
\end{equation}
Conserved variables, {i.e.}, density and momentum are defined as moments of the discrete distribution function:
\begin{equation}
    \rho = \sum_\alpha f_\alpha,
\end{equation}
\begin{equation}
    \rho \bm{u} = \sum_\alpha \bm{c}_\alpha f_\alpha.
\end{equation}

\subsubsection{\gls{lbm} for phase-field equation}

The phase-field equation is modeled using a modified \gls{lb} scheme implemented as~\cite{walsh2010macroscale,cartalade2016lattice}:
\begin{multline}
  a_s^2(\bm{n}) h_\alpha(\bm{x} + \bm{c}_\alpha \delta x, t + \delta t) = h_\alpha(\bm{x},t) \\
  -\left( 1 - a_s^2(\bm{n})  \right ) h_\alpha(\bm{x} + \bm{c}_\alpha \delta x, t) - \\ 
   \frac{1}{\eta_\phi (\bm{x},t) }
  \left [ h_\alpha(\bm{x},t) - h_\alpha^{eq}(\bm{x},t) \right]  + w_\alpha Q_\phi (\bm{x},t)\frac{\delta t}{\tau_0},
  \label{d}
\end{multline}
where the scalar function $Q_\phi$ is the source term of the phase-field defined as:
\begin{equation}
    Q_\alpha = (\phi - \phi^3) + \lambda (U + \theta) (1 - \phi^2)^2,
\end{equation}
while the \gls{edf} $h_\alpha^{eq}$ is defined as:
\begin{equation}
    h_\alpha^{eq} = w_\alpha \left( \phi - \frac{1}{c_s^2} \bm{c}_\alpha \cdot \frac{W_0^2}{\tau_0} |\bm{\nabla} \phi|^2 \frac{\partial (a_s(\bm{n})^2)}{\partial \bm{\nabla} \phi} \frac{\delta t}{\delta x} \right).
    \label{e}
\end{equation}
The local value of the order parameter $\phi$ is computed as:
\begin{equation}
    \phi = \sum_{\alpha} h_\alpha,
\end{equation}
while the relaxation is set to:
\begin{equation}
  \eta_\phi = \frac{1}{c_s^2}a_s^2(\bm{n})\frac{W_0^2}{\tau_0} + \frac{\delta t}{2}.
\end{equation}

\subsubsection{Finite-difference solver for species and energy equations}
Balance equations for supersaturation and temperature are solved using a finite-difference scheme with a simple first-order time-stepping coupled to a fourth-order central discretization in space for diffusion terms and a third-order weighted essentially non-oscillatory (WENO) approximation for convective terms~\cite{liu1994weighted}. Related researches are investigated from~\cite{liu1994weighted,cockburn1998essentially,shu2020essentially}.
\subsection{Evaluation of thermo-physical properties}

The physical parameters of the pure (S)-mandelic acid at temperature $T$ = 298.15K (or 25 $^{\circ}$C) are listed in Table \ref{phy}. The enthalpy of crystallisation $\Delta H_{cryst}$ characterizes the energy difference per mole of S-ma between the solid and liquid(melt) phase and is represented here as negative value of the enthalpy of fusion~\cite{emel2018experimental}. The value of specific heat capacity $c_{p,L}$ is that of water (being by far the dominating component) and $c_{p,S}$ is for racemic MA~\cite{sapoundjiev2005determination}. Furthermore, $\kappa_S$ denotes the thermal diffusivity of the crystal and $\kappa_L$ represents the thermal diffusivity of water~\cite{slack1979thermal,speedy1982stability}, while $D$ is the diffusion coefficient for the solution~\cite{tanner1983intracellular}.
 \begin{table}[!htbp]
 \centering
 \setlength{\tabcolsep}{1.4mm}{
 \caption{Physical parameters used for modeling single S-ma crystal growth at temperature of 25$^{\circ}$C.}
\begin{tabular}{cccc}
\hline
Property & Value  &  Unit & Ref.\\
 \hline
Enthalpy of crystallisation $\Delta H_{cryst}$        &   -18.5  &  kJ/mol  &  \cite{emel2018experimental}    \\
Specific heat capacity for solid $c_{p,S}$    &   160.5   &   J/mol $\cdot$ K   &  \cite{sapoundjiev2005determination} \\
Specific heat capacity for liquid $c_{p,L}$    &   75   &   J/mol $\cdot$ K   &  \cite{sapoundjiev2005determination} \\
Thermal diffusivity for solid $\kappa_S$    &   1.1     &   mm$^2$/s  &  \cite{slack1979thermal} \\
Thermal diffusivity for liquid $\kappa_L$   &   0.146   &   mm$^2$/s  & \cite{speedy1982stability} \\
Diffusion coefficient in liquid $D$       &   1.2$\times 10^{-3}$   &   mm$^2$/s   &  \cite{tanner1983intracellular}\\
Crystal growth rate constant $k_0$       &   1.0$\times 10^{-5}$   &   cm/s   &  \cite{zhang2010nucleation}\\
Density of solid $\rho_S$      &  1.341   &  $\mathrm{g/cm^3}$ &  \cite{densityma}\\
Density of fluid $\rho_L$      &  1.0     &  $\mathrm{g/cm^3}$ &  \cite{patterson1994measurement}\\
\hline  
\label{phy}
\end{tabular}}
\end{table}

\section{\label{sec:level3}Experimental setup}

\begin{figure}[!ht]           
\centering	
	\includegraphics[width=0.8\textwidth]{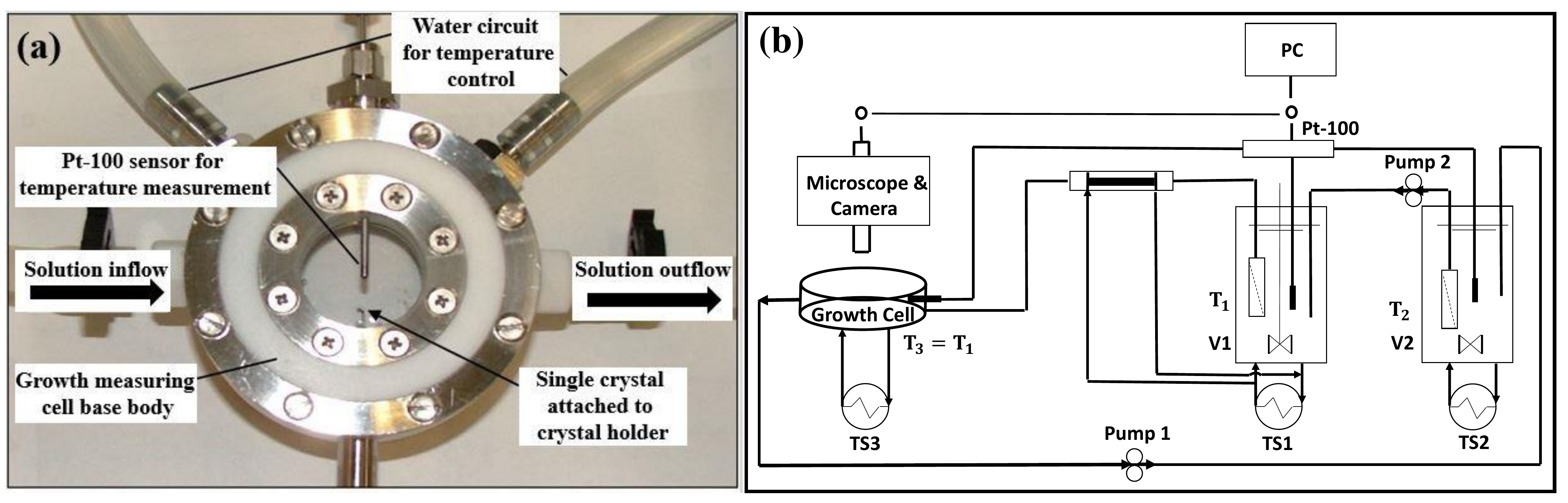}	
	\caption{Single-crystal growth cell used for all experiments: (a) photograph; (b) Schematic diagram of experimental arrangement for the measurement of a single crystal growth rates~\cite{gou2012investigation, Juan}}.
	\label{ba}
\end{figure}

All experimental data for the single S-ma crystal growth rate in the growth cell have been obtained from~\cite{gou2012investigation, Juan}. The corresponding experimental setup is illustrated in Fig.~\ref{ba}. A supersaturated aqueous solution of mandelic acid is pumped into a constant-temperature cylindrical crystallization cell, with solution temperatures varying between 20 and 30$^{\circ}$C. The temperature within the cell is maintained constant via a water-based cooling/heating system connected to a Pt-100 sensor monitoring the temperature at the center of the cell. Vessel 2, denoted V2 in Fig.~\ref{ba}b contains a saturated solution at temperature $T_2$ while vessel 1 (V1) was set to a lower temperature $T_1$, corresponding to the temperature of the cell. To create the supersaturated solution, the initially saturated solution in V2 is pumped into V1 and cooled down to $T_1$ before entering the growth cell. This effectively allows to control the supersaturation level of the incoming solution by choosing temperature $T_1$. Based on the solutions in the two vessels, the normalized supersaturation is defined as~\cite{mullin2001crystallization}:
\begin{equation}
    U = \frac{C_{sat,2} - C_{sat,1}}{C_{sat,1}}
    \label{U}
\end{equation}
To start the experiment, the supersaturated solution is continuously pumped from vessel 1 to the growth cell, in which a single (S)-mandelic crystal is glued on the pin head of a crystal holder. Then, the solution is recycled to vessel 2 and the concentration of the solution is compensated. In that way, a stable degree of supersaturation is guaranteed during the whole process. A microscope with camera (Stemi2000C, Carl Zeiss Co.) is used to take pictures of the single crystal at every one hour. The images are afterwards post-processed by applying Carl Zeiss' Axio Vision software~\cite{gou2012investigation}.\\
In the experimental setup, small temperature differences and gradients cannot be measured, since this quantity is measured at a single point. Due this fact it is attractive to analyze, the temperature field within the entire growth cell numerically.\\ 
\section{\label{sec:IV}Simulations and analysis of the results}
\subsection{\label{sec:A}Validation of the finite-difference (FD) solver}
In the present study, the FD method is adopted to solve for energy and species due to the large ratio between the value of thermal diffusivity and of mass diffusion coefficient; the corresponding Lewis number (the ratio between thermal diffusivity and mass diffusion) is of the order of $10^2-10^3$. Finite differences are more suitable for low values of the diffusion coefficient, since \gls{lbm} becomes numerically unstable at very low non-dimensional diffusion coefficients, the relaxation time $\tau$ becoming close to 0.5.
\subsubsection{\label{sec:A1}Self-convergence of FD method}
\begin{figure}[!ht]                          
	\centering
	\includegraphics[width=0.5\textwidth]{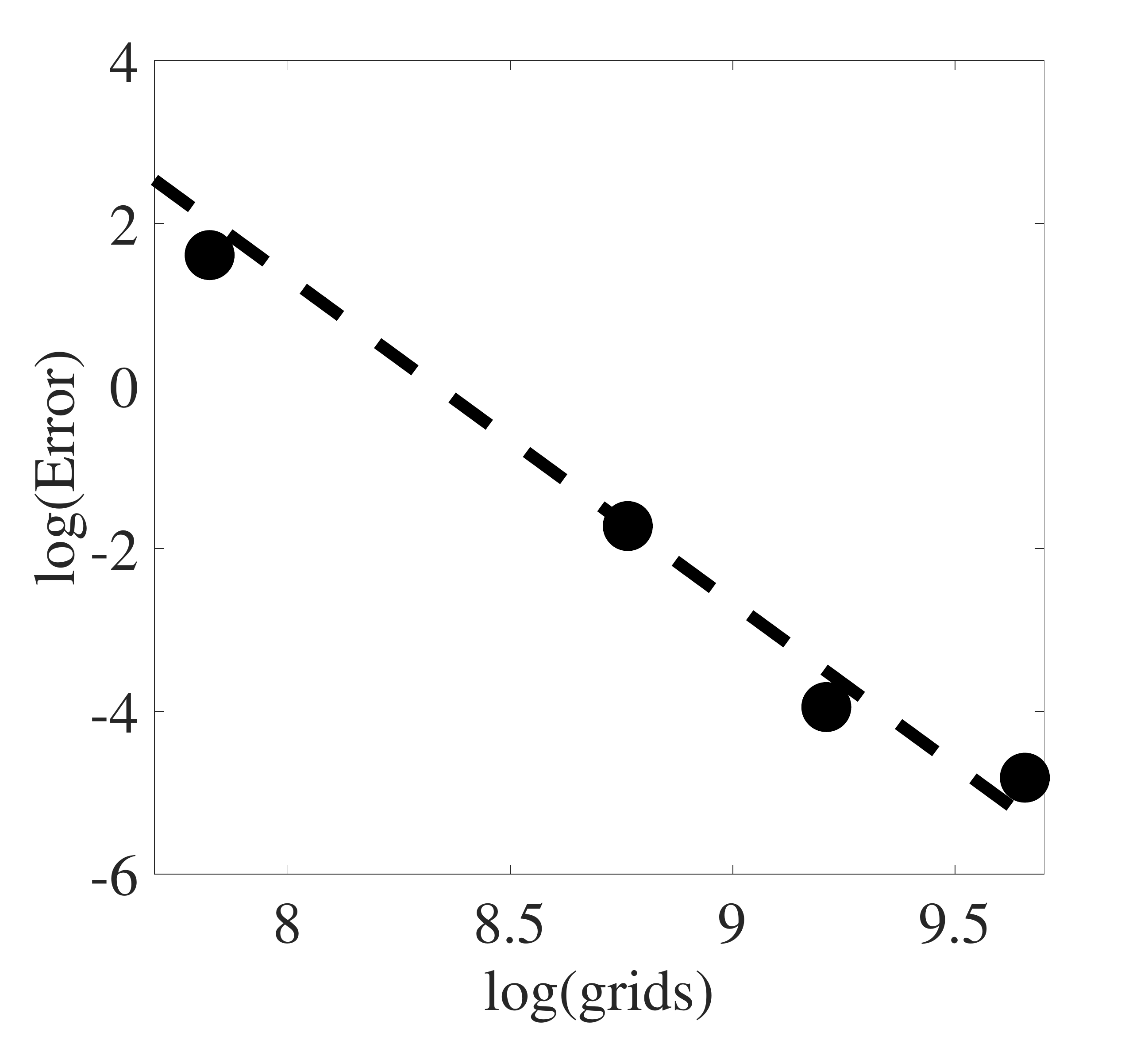}		
	\caption{Scaling of the $l^2$ error norm as obtained from the grid convergence study. Black markers represent error data from the simulations (see Table \ref{t1}) while the black dashed line displays the theoretical -4 slope.}
	\label{11}
\end{figure}

As known, \gls{lbm} is of second-order accuracy for the phase-field model~\cite{tan2022mandelic}. Here, the accuracy of the finite-difference method is checked by considering only diffusion for a case with non-homogeneous concentrations. In this test-case, periodic boundaries are implemented in a 2D box of size $[-1,1] \times [-1,1]$mm.\\
The concentration profile is set as a Gaussian hill following \cite{kruger2017lattice,fedi2010new}:
\begin{equation}
 C(\mathbf{x}, t) =\frac{\Psi_0}{2 \pi \sqrt{|\bm{\sigma_t}|}}  {\rm exp} \left( -\frac{1}{2} \bm{\sigma_t}^{-1}: \mathbf{x}^2 \right).
    \label{ini}
\end{equation}
where $\Psi_0 = 2 \pi \sigma^2_0$ with initial variance $\sigma_0 = 0.01$mm. The tensor $\bm{\sigma_t} = \sigma_0^2 \textbf{I} + 2t \textbf{D}$, $|\bm{\sigma_t}|$ are the determinant value and $\bm{\sigma_t}^{-1}$ is inverse matrix of $\bm{\sigma_t}$, respectively. Quantity $\mathbf{I}$ is the unit matrix. Note that $\sigma_0$ is small enough in the present case, so that periodic boundary conditions are suitable. \\
The simulations are conducted using four different spatial resolutions, $\delta x\in\{0.04, 0.025, 0.02, 0.016\}$mm. Since the overall size of the numerical domain is kept fixed, an improved spatial resolution automatically comes with a larger number of grid points. Then, the results are compared with the analytical solution (see Eq.(\ref{ini})) at time $t = 10$s.\\
The $\mathit{l^2}$ relative error norm is calculated based on the concentration profiles over the entire domain. The $\mathit{l^2}$ norm is defined as:
\begin{equation}
{\rm E}_{\mathit{l^2}}=\sqrt{\frac{\sum_i \left( C_{i} - C_{an,i} \right) ^2}{\sum_i C_{an,i}^2}}
\label{er}
\end{equation}
where $C_i$ represents the concentration obtained numerically at a certain position in the box and $C_{an}$ denotes the analytical solution from Eq.~(\ref{ini}). The errors obtained from the different simulations are illustrated in figure~\ref{11}.

\begin{table}[!htbp]
\centering
\setlength{\tabcolsep}{2.5mm}{
\caption{Relative $\mathit{l^2}$ errors of the scalar variable $C$ for different resolutions}\label{t1}
\begin{tabular}{|c|c|c|c|c|}
\hline
Numerical grid  &  $50 \times 50$  &  $80 \times 80$  & $100 \times 100$  &  $125 \times 125$\\
\hline
$E_{l^2}$ &   5.0565  &  0.1787  &  0.0193  &  0.0081 \\
\hline
\end{tabular}}
\end{table}

As observed from this plot, the numerical scheme is convergent as the error decreases with resolution. Furthermore, as expected from theoretical analyses, a fourth-order convergence is obtained for the finite-difference solver.

\subsubsection{\label{sec:A3}Limiting case: adiabatic single-crystal growth cell}
In this section, in order to get rough insight regarding the temperature range possible, an adiabatic single-crystal growth cell is computed using mass and energy conservation for the purpose of verifying the proper function of the hybrid \gls{lbm}/FD solver for describing single S-ma crystal growth rate and temperature within the cell.

\begin{figure}[!ht]                        
\centering	
	\includegraphics[width=0.4\textwidth]{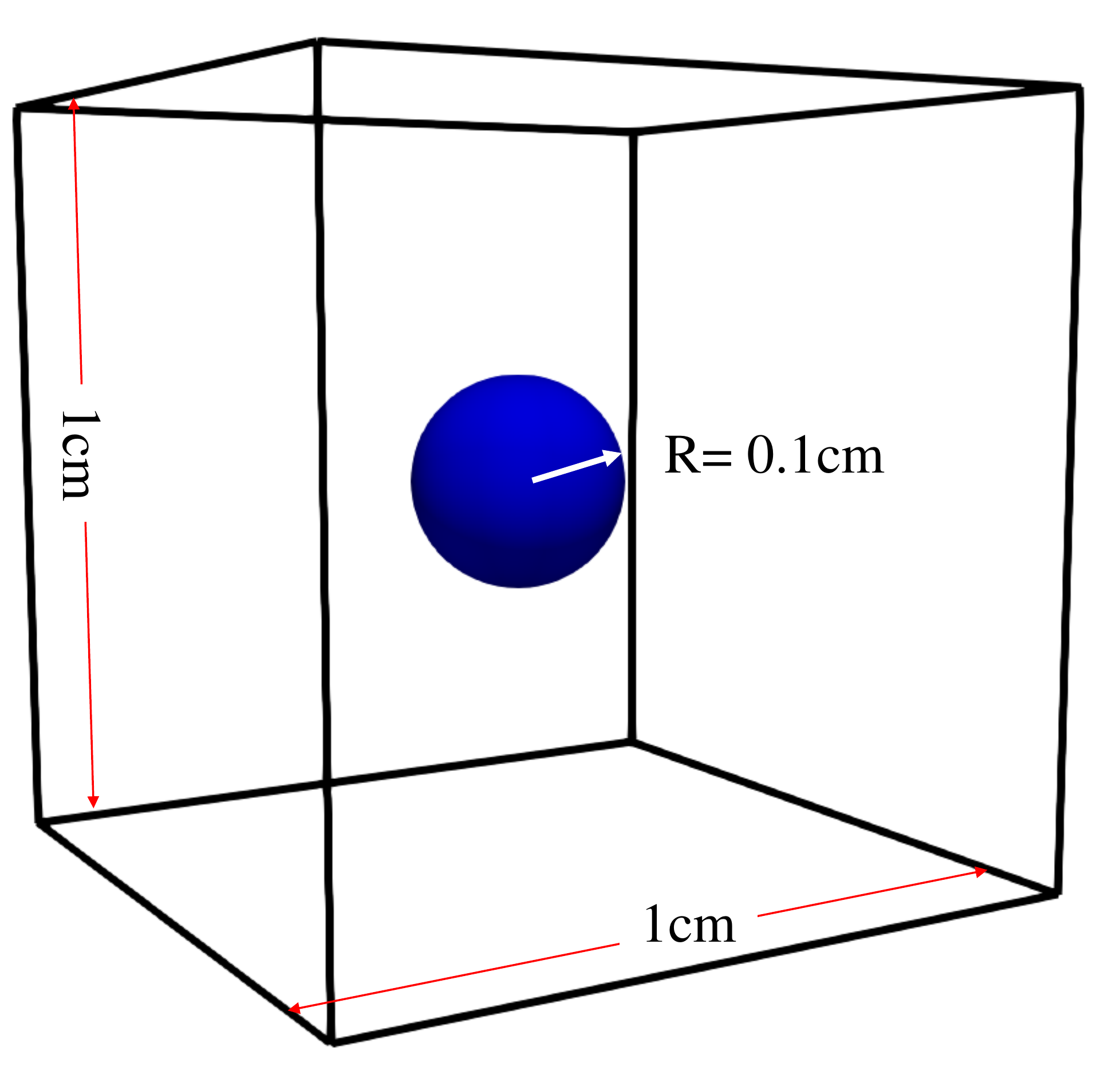}
	\caption{Schematic of adiabatic cell box in 3D.}
	\label{adia}
\end{figure}

The model is based on a square box in 3D (see Fig.~\ref{adia}) with a side length of 1cm. The seed is set in the center of the box with initial radius $R$ = 0.1cm. The initial concentration of the supersaturated aqueous S-ma aqueous solution is 0.887 mmol/cm$^3$ in this closed adiabatic system. The S-ma crystal keeps growing until the solution concentration reaches equilibrium.
The liquid phase mass balance involving liquid phase concentration $c$ reads:
\begin{equation}
    V_L \frac{dc}{dt} = -k(T) A_{solid}(c - c_{sat}(T)) = -k(T)4 \pi R^2(c-c_{sat}(T))
\end{equation}
where $V_L$ is the volume of the adiabatic box (here, 1cm$^3$); variable $A_{solid} = 4\pi R^2$ is the surface area of the solid; $V_{solid} = \frac{4}{3} \pi R^3$ is its volume; $c_{sat}(T)$ is the saturation concentration (see Eq.(\ref{sat})) and $k(T)$ the growth rate constant (see Eq.(\ref{kinetic})), both at temperature T.
The ordinary differential equation describing the liquid phase concentration $c$ is:
\begin{equation}
   \frac{dc}{dt} = - \frac{k(T)}{V_L} 4 \pi R^2 (c - c_{sat}(T)) 
\end{equation}
The solid phase mass balance reads (assuming that the density does not depend on temperature, since temperature differences are expected to be low):
 \begin{equation}
     \rho_S \frac{d V_S}{dt} = \rho_S 4 \pi R^2 \frac{dR}{dt} = k(T)4 \pi R^2 (c - c_{sat}(T))
 \end{equation}
 where $\rho_S$ is the density of the mandelic acid crystal. The second ordinary differential equation describing radius is:
 \begin{equation}
     \frac{dR}{dt} = \frac{k(T)}{\rho_S}(c - c_{sat}(T)) 
 \end{equation}

The third ordinary differential equation representing energy (here in the form of temperature) is:
\begin{equation}
    \frac{dT}{dt} = - \frac{\Delta H_{cryst}}{V_L \rho_L c_{p,L} + \frac{4}{3} \pi R^3 \rho_S c_{p,S}} k(T) 4 \pi R^2 (c - c_{sat}(T))
\end{equation}
The value of specific heat capacity $c_{p,L}$ is that of water, and $c_{p,S}$ is from the S-mandelic acid, taken from~\cite{sapoundjiev2005determination}. The saturation function for a mandelic acid aqueous solution is~\cite{lorenz2002enantiomeric}:
\begin{equation}\label{sat}
    c_{sat}(T) = -0.005006 + 0.00001923T
\end{equation}
where temperature $T$ with unit K and the kinetic growth rate constant is:
\begin{equation}\label{kinetic}
    k(T) = k_0 e^{-E/(RT)}
\end{equation}
where $k_0$ is the crystal growth rate constant with the unit [cm/s]; $E$ is the activation energy, with unit [J/mol]. Due to the small temperature range covered, the growth rate coefficient $k(T)$ was assumed to be constant in the temperature range between 20 and 30 $^\circ$C. The value used is given in Table~\ref{phy}.

\begin{figure}[!ht]                        
\centering	
	\includegraphics[width=0.5\textwidth]{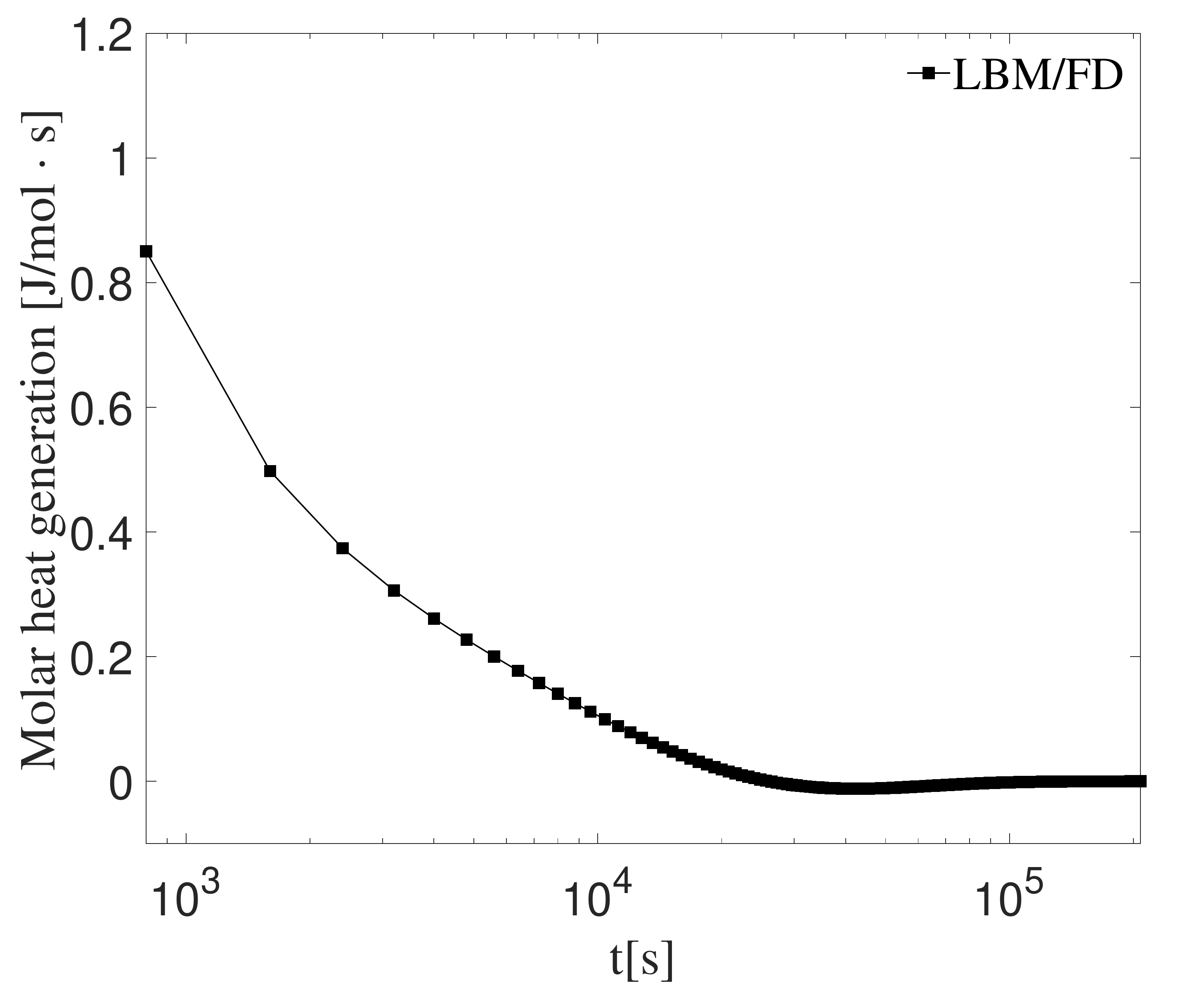}
	\caption{Evolution of average molar heat generation with time connected to spherical crystal growth.}
	\label{heattime}
\end{figure}

In the numerical simulation based on the hybrid \GLS{lbm}/FD solver, the spatial discretization is 0.01cm (leading to a grid [100 $\times$ 100 $\times$ 100]) and the time-step is 0.005s. The physical parameters of S-ma are selected based on Table \ref{phy}.
\begin{figure}[!ht]                        
\centering	
	\includegraphics[width=0.8\textwidth]{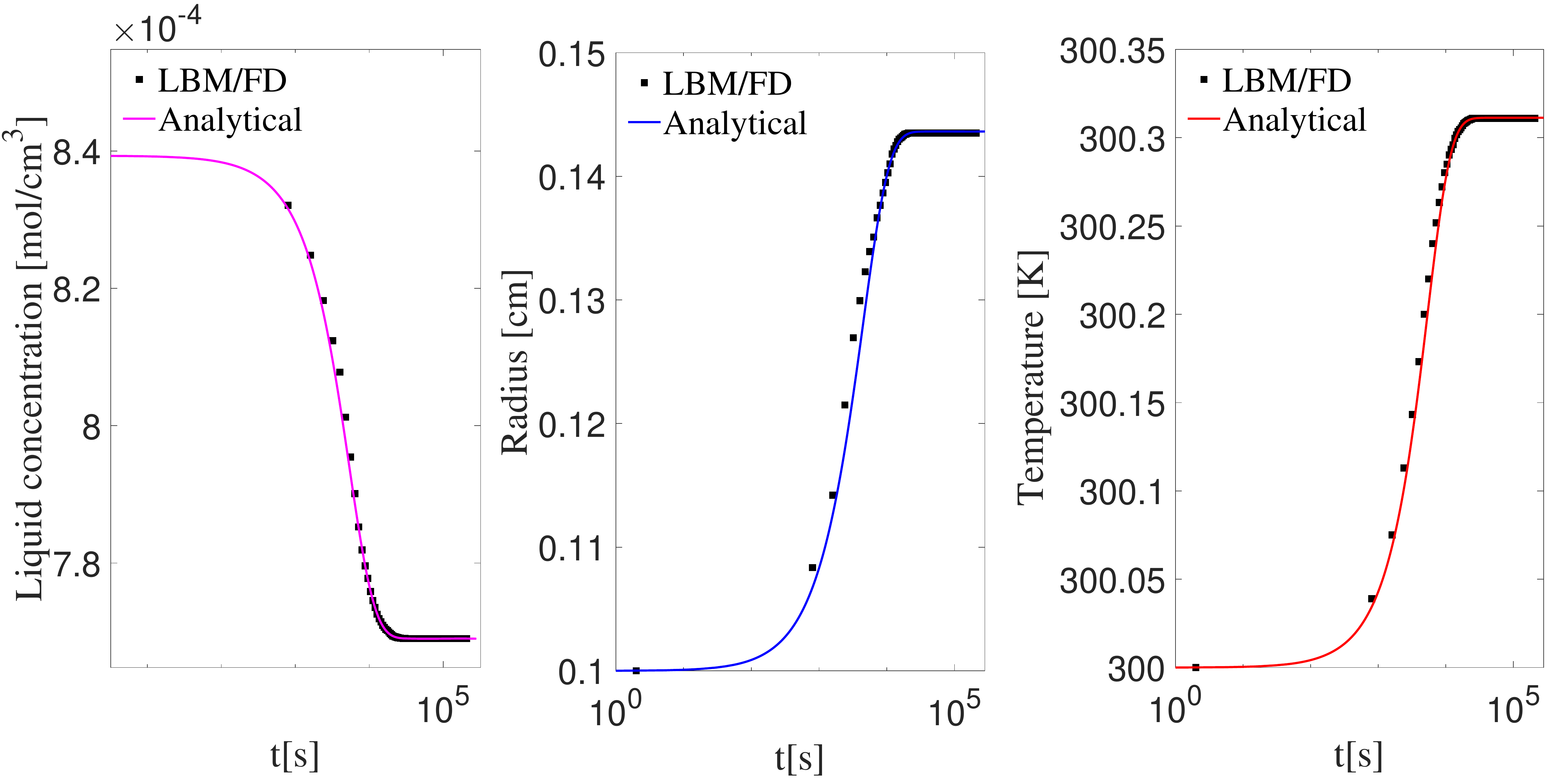}
	\caption{Plots showing average concentration (left), crystal radius (center) and average temperature (right) as function of time as obtained from the hybrid \gls{lbm}/FD solver, compared with analytical solution. Note the horizontal logarithmic scale due to the long duration of the process.}
	\label{mat}
\end{figure}
Figure~\ref{mat} shows average concentration, crystal radius, and average temperature as function of time. It can be observed that the numerical solution obtained with the hybrid \gls{lbm}/FD solver matches well with the analytical solutions derived from mass and energy conservation. 
\begin{figure}[!ht]                        
\centering	
	\includegraphics[width=0.7\textwidth]{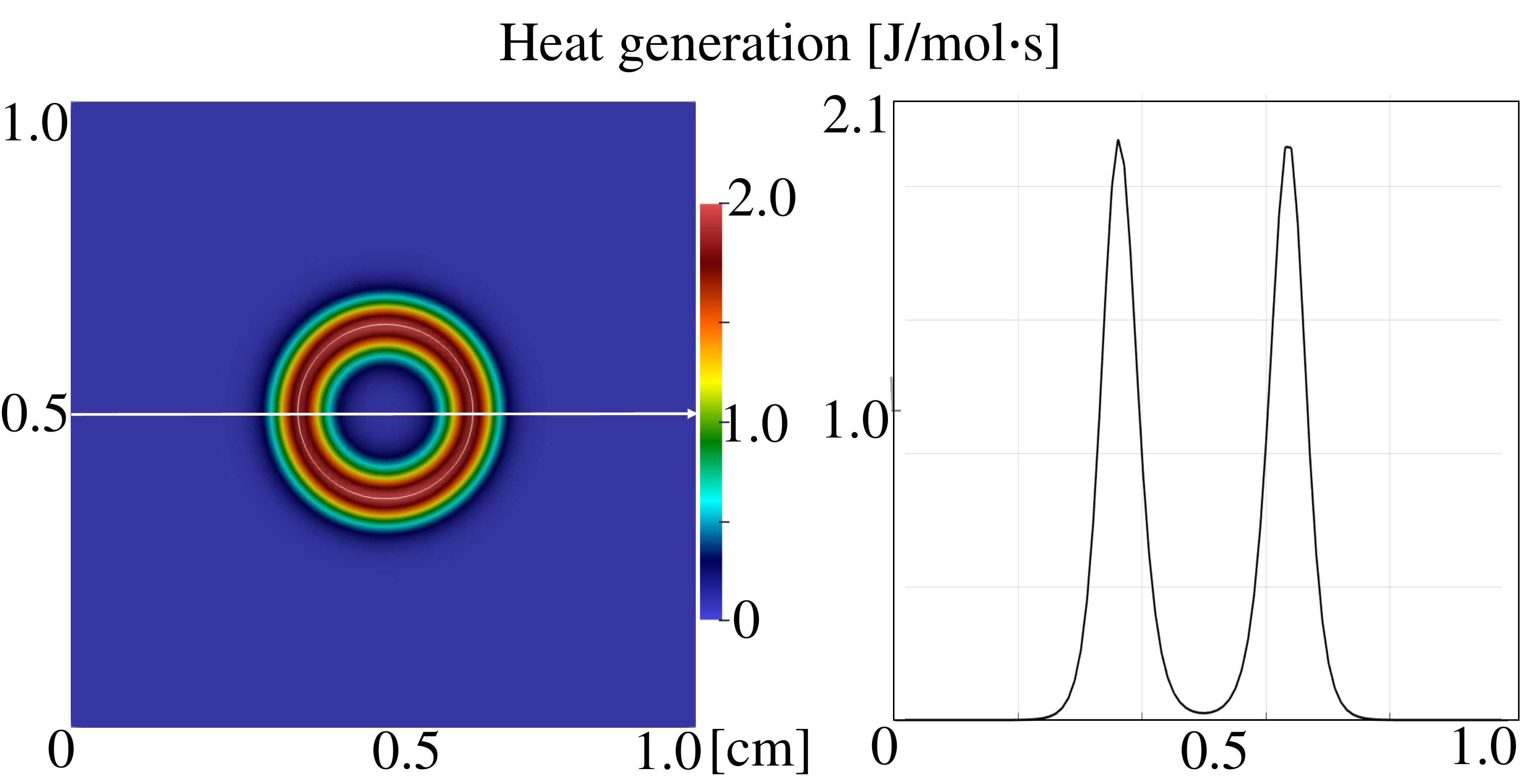}
	\caption{Molar heat generation during growth of spherical crystal at time $t = 1$ hour.}
	\label{heatcell}
\end{figure}
Figures~\ref{heatcell} and~\ref{heattime} illustrate the heat release process during phase-change from the liquid into the solid state at the interface of the crystal. Eventually, the heat release is stopped when the crystal stops growing (equilibrium state has been reached).
This successfully terminates the verification procedure for the developed numerical model. It can now safely be used to investigate growth rates and possible temperature effects for a single S-ma crystal.
\subsection{\label{sec:B}Validation for S-ma crystal growth including temperature effects}
\subsubsection{\label{sec:B1}Experiments vs. numerical simulation for different temperatures}
\begin{figure}[!ht]                        
\centering	
	\includegraphics[width=0.5\textwidth]{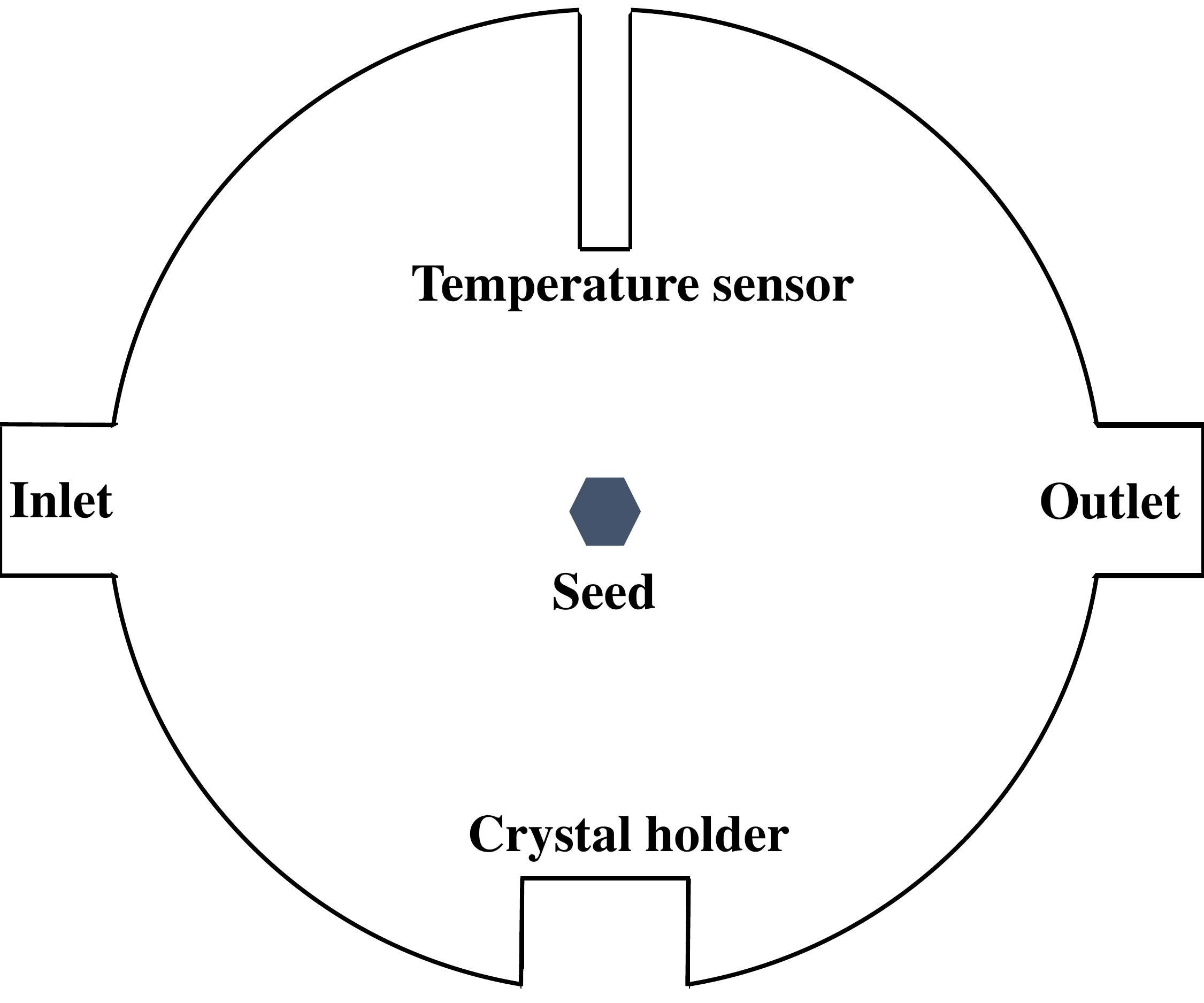}
	\caption{Reactor geometry employed for all 2D simulations.}
	\label{2dgeometry}
\end{figure}

An excellent agreement between numerical predictions and experimental observations was observed in a previous study when neglecting the influence of changes in temperature \cite{tan2022mandelic}. In order to check now the ability of the solver to correctly describe S-ma crystal growth at different temperatures (from 20$^\circ$C to 30$^\circ$C), 2D simulations are carried out using the real reactor geometry. The reduction to two dimensions is justified by the fact that, in all conditions considered here, the crystal follows a platelet growth mode leading to a clear separation of scales between growth in axial or in planar directions, ensuring also symmetry of the flow field~\cite{Henniges2017}. The 2D geometry used for the simulations is shown in Fig.~\ref{2dgeometry}. First, configurations are considered where forced convection is negligible. For all experiments presented in this section the initial seed is a hexagonal crystal. The initial supersaturation is $U=0.045$, and the temperature is $T = 20^\circ$C, $T=25^\circ$C, or $T=30^\circ$C, respectively. The employed physical parameters have been given in Table \ref{phy}. All simulations are carried out with a spatial resolution of $\delta x=0.1$~mm. The interface thickness is set to $W_0 = 0.25$mm, the relaxation time to $\tau_0 = 0.02$s, and the coupling coefficient $\lambda = 3$ was chosen as a standard value for the phase-field method for dendrite growth~\cite{ramirez2004phase}. At the walls of the reactor, zero-flux boundary conditions are applied to both the species and phase fields. A constant wall temperature (set as the value of $T_1$, see figure~\ref{ba}) is used as boundary condition for the energy equation. At the inlet a constant supersaturation is imposed, following the implementation described in~\cite{kruger2017lattice} for the boundary condition.

\begin{figure}[!ht]                        
\centering	
	\includegraphics[width=0.3\textwidth]{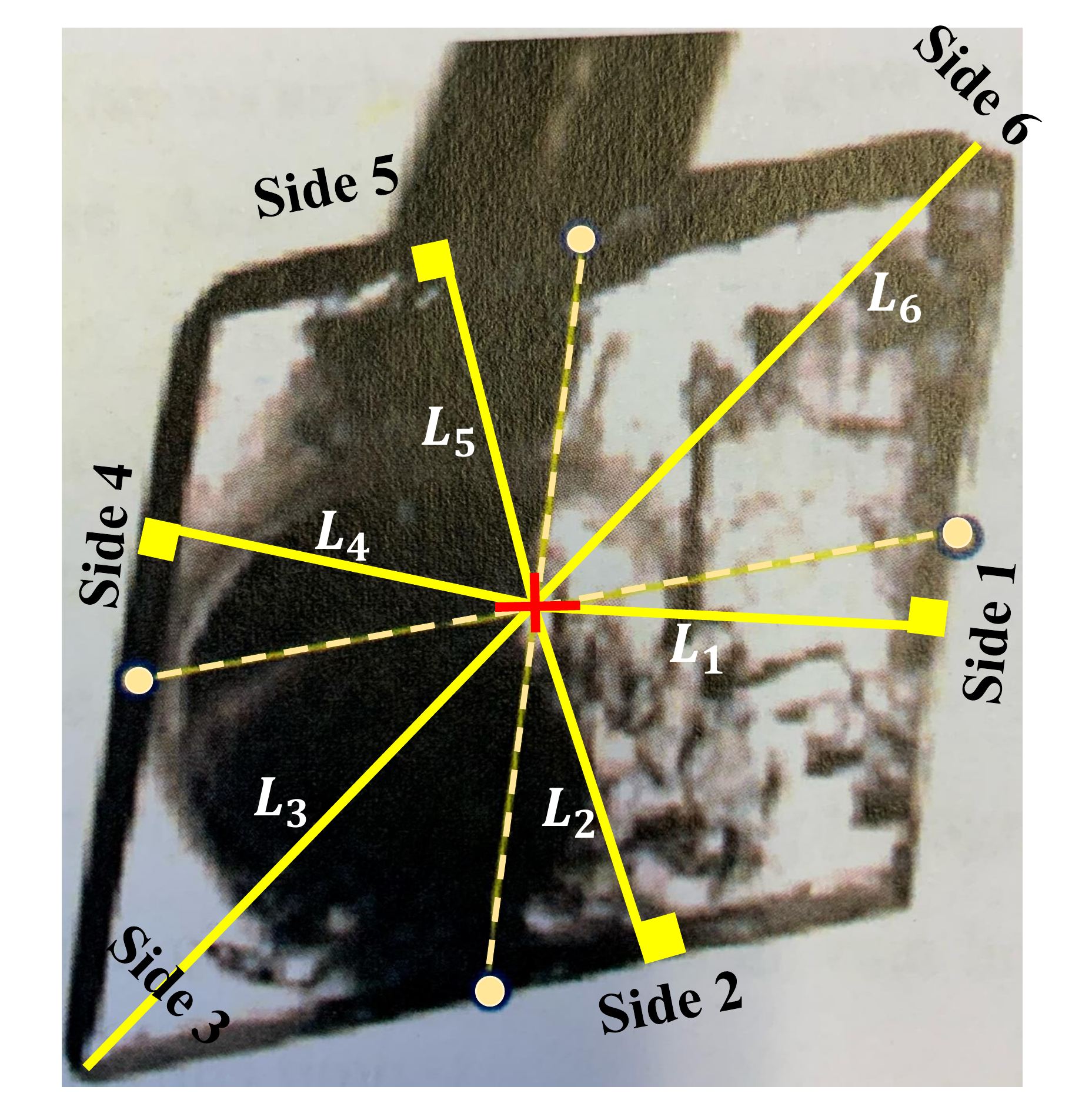}
	\caption{Method used to number the crystal sides and the associated normal directions~\cite{Juan}.}
	\label{side}
\end{figure}

\begin{table}[!htbp]
\centering
\setlength{\tabcolsep}{1.4mm}{
\caption{Comparison between experiments and simulations for initial supersaturation $U = 0.045$ as a function of temperature~\cite{Linzhu}.}\label{asdd}
\begin{tabular}{|c|c|c|c|c|}
\hline
Average growth rate [mm/h] & $T=20^\circ$C & $T=25^\circ$C & $T=30^\circ$C\\
\hline
Experiments &   0.011 &  0.023 & 0.0321   \\
\hline
Simulations   &   0.0096   &   0.0222  &  0.0317 \\
\hline
\end{tabular}}
\end{table}

In Table \ref{asdd}, $G_{th}$ is the average growth rate (in mm/h) obtained as $G_{th} = (L_1 + L_2 + L_3 + L_4 + L_5 + L_6)/6t$ in both experiment and simulation. The comparison between computed and measured values points to a good agreement at all temperatures. $L$ is the normal length from the center to every side of the crystal(see Fig.~\ref{side}). It is observed that the S-ma growth rate increases with temperature. As far as can be judged from only 3 values (no other conditions have been investigated experimentally), a quite linear behavior is observed between temperature and growth rate in the range studied. It is now interesting to check the occurrence and strength of possible temperature gradients within this growth cell.

\subsubsection{\label{sec:B2}Occurrence of temperature gradients during crystal growth}
\begin{figure}[!ht]                          
	\centering
	\includegraphics[width=0.6\textwidth]{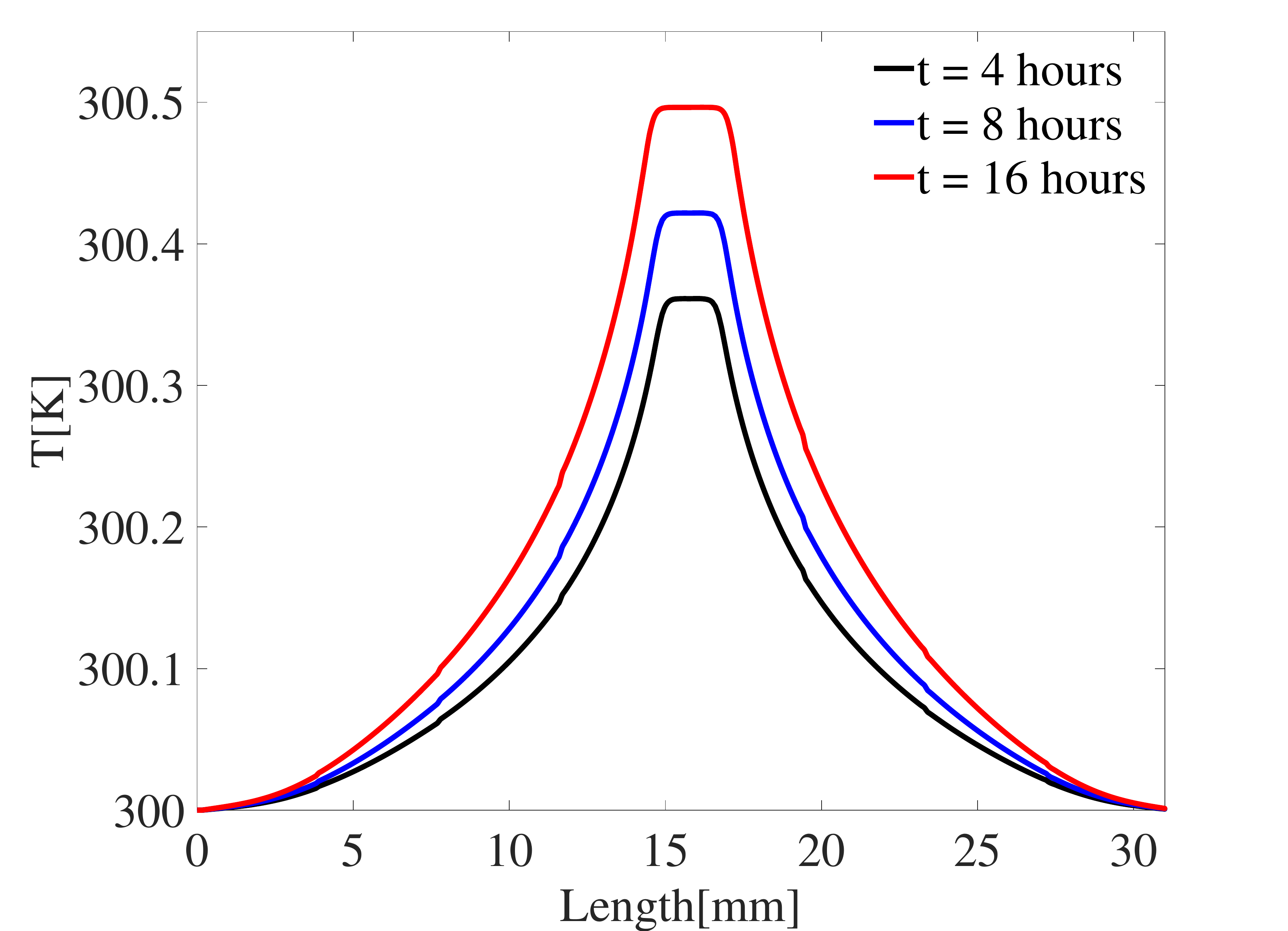}	
	\caption{Temperature profile around the center-line of the numerical domain at times $t=$4, 8 and 16 hours, respectively.}
	\label{temperaturep}
\end{figure}

Although the temperature of the single-crystal growth cell in the experimental setting is kept constant through the walls at the temperature of vessel V1, the growth of the crystal generates heat at the interface between liquid and solid phase. The Pt-100 sensor used for the temperature measurements in the experiment delivers only a point value and cannot be used to track possible gradients. Hence, in this section, the temperature field in the whole growth cell is studied numerically. The initial supersaturation is kept at $U_0 = 0.045$ as in the previous section.\\
\begin{figure}[!ht]                          
	\centering
	\includegraphics[width=0.6\textwidth]{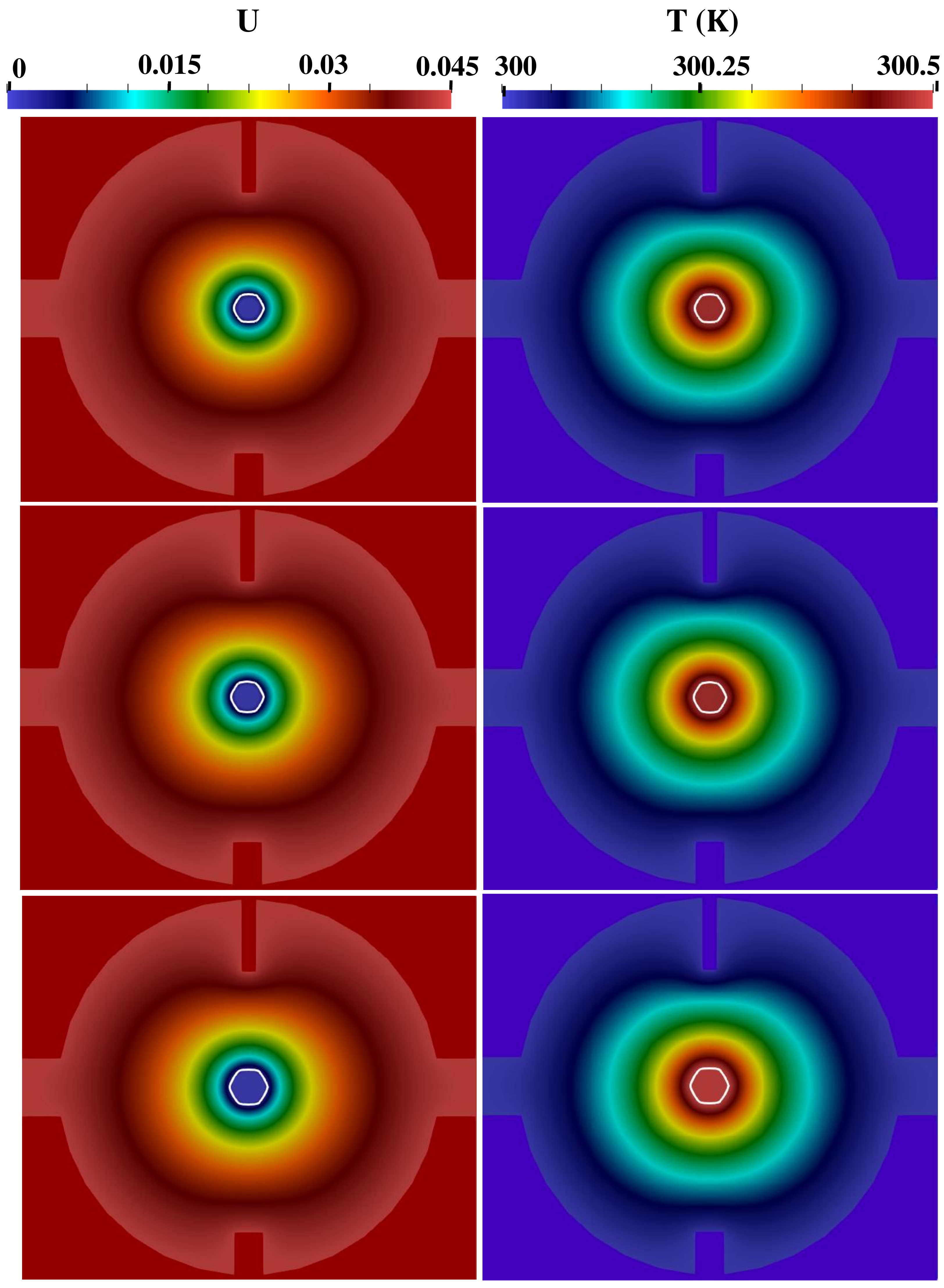}	
	\caption{Instantaneous supersaturation and temperature fields in the growth cell for initial supersaturation $U_0 = 0.045$ at times $t = 4$ hours (top), 8 hours (center), 16 hours (bottom), respectively.}
	\label{temperature}
\end{figure}
In Figure \ref{temperature}, it is seen that the highest temperature in the crystal as well as the solution temperature far from the crystal are still found at around 300K. However, a maximum difference in temperature of the order of $0.5^{\circ}$C is indeed observed within the cell, with a maximum temperature close to the interface. Though small, this shows that temperature differences do exist within the single-crystal growth cell. Since these differences appear locally, non-negligible temperature gradients will occur as well. Figure \ref{temperaturep} demonstrates that the temperature increase to has a peak value in the range of the single crystal and decreases in the fluid phase to the wall of the cell.

\begin{figure}[!ht]                          
	\centering
	\includegraphics[width=0.8\textwidth]{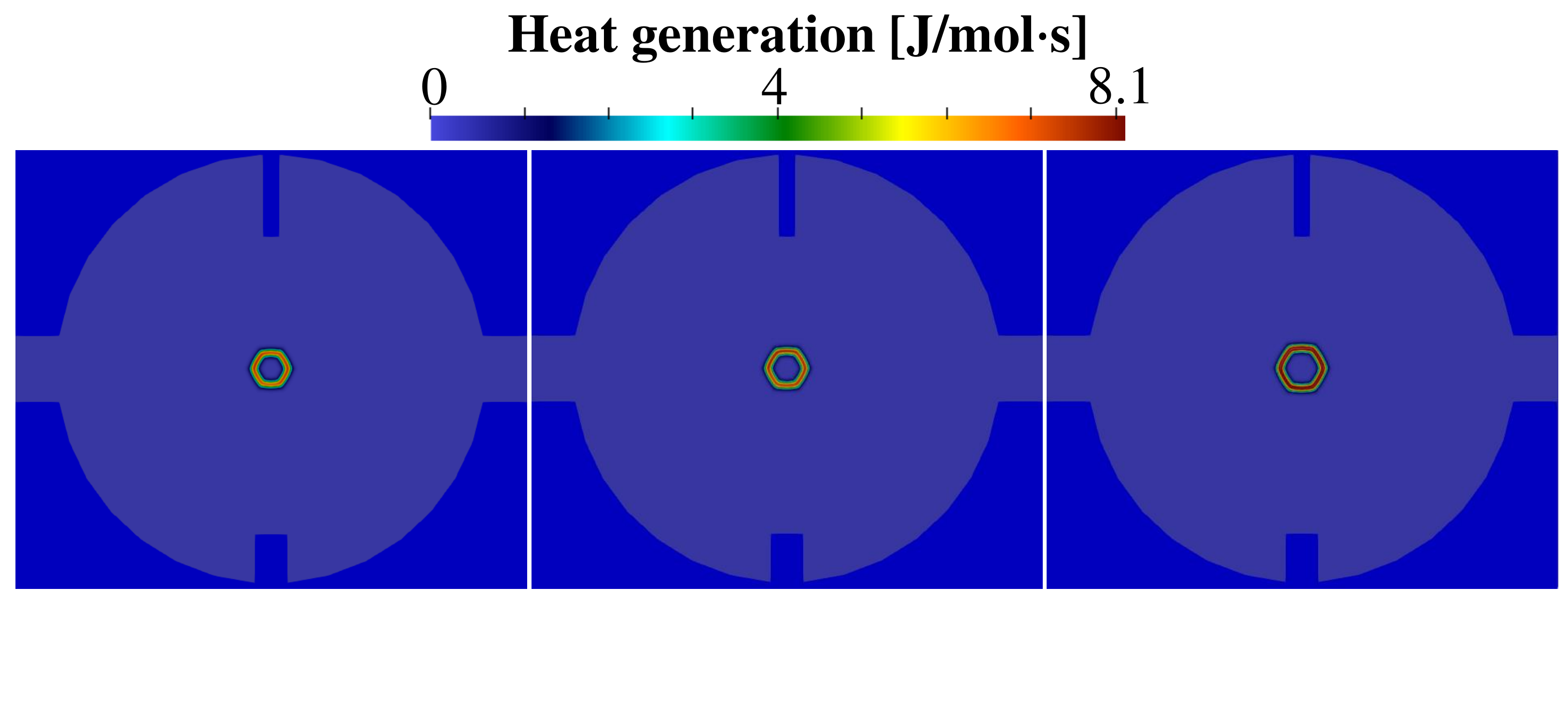}	
	\caption{Instantaneous heat generation at the interface of the crystal at times $t=$4, 8 and 16 hours (from left to right), respectively.}
	\label{heat}
\end{figure}

\begin{figure}[!ht]                          
	\centering
	\includegraphics[width=0.6\textwidth]{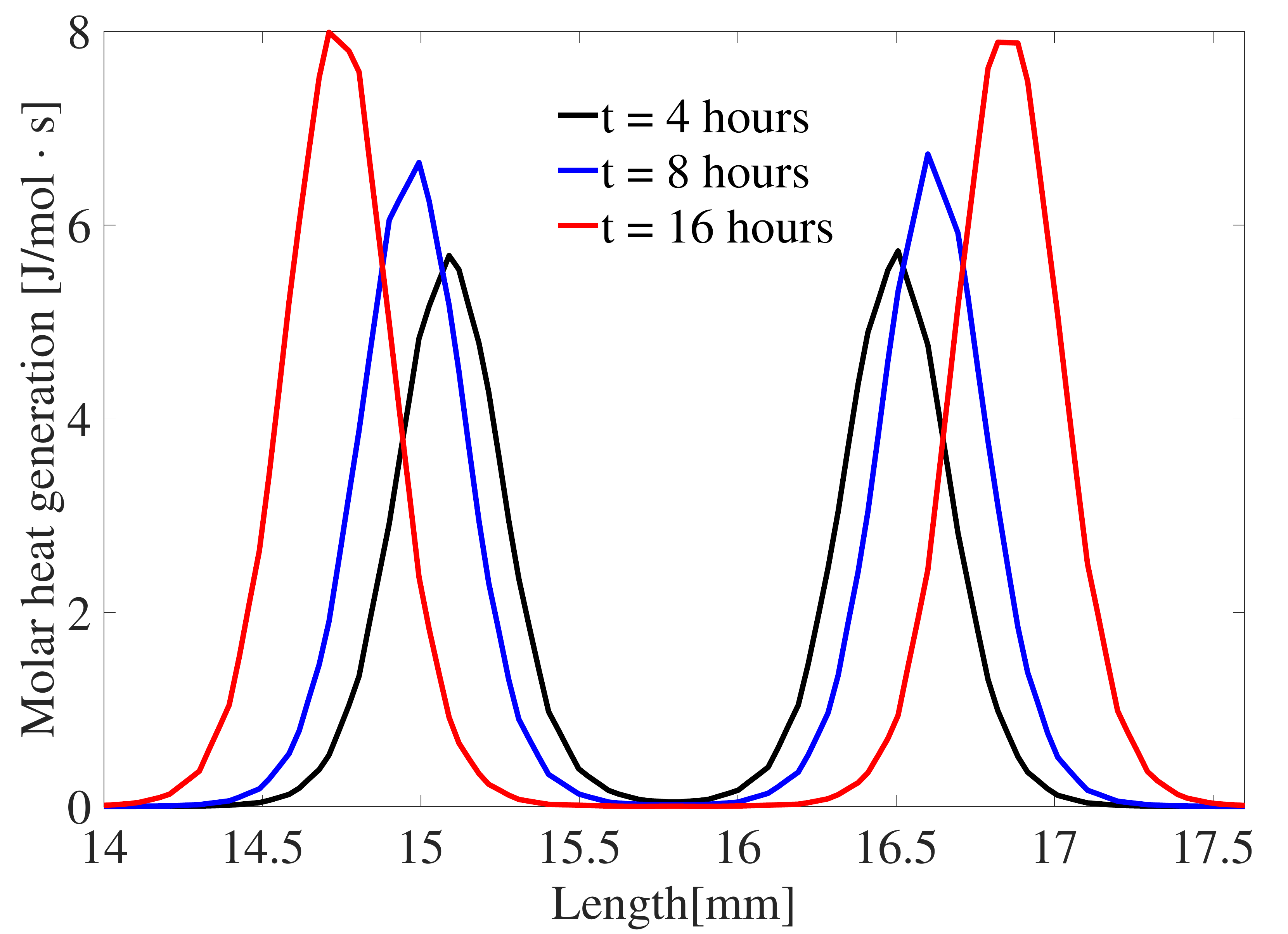}	
	\caption{Instantaneous heat generation along the centerline of the numerical domain at times $t=$4, 8 and 16 hours, respectively.}
	\label{heats2}
\end{figure}

Figure \ref{heat} and \ref{heats2} show the heat release process during phase-change from the liquid into the solid state at the interface of the crystal.


\subsubsection{\label{sec:B3}Ventilation and temperature effects during S-ma crystal growth}
\begin{figure}[!ht]                          
	\centering
	\includegraphics[width=0.8\textwidth]{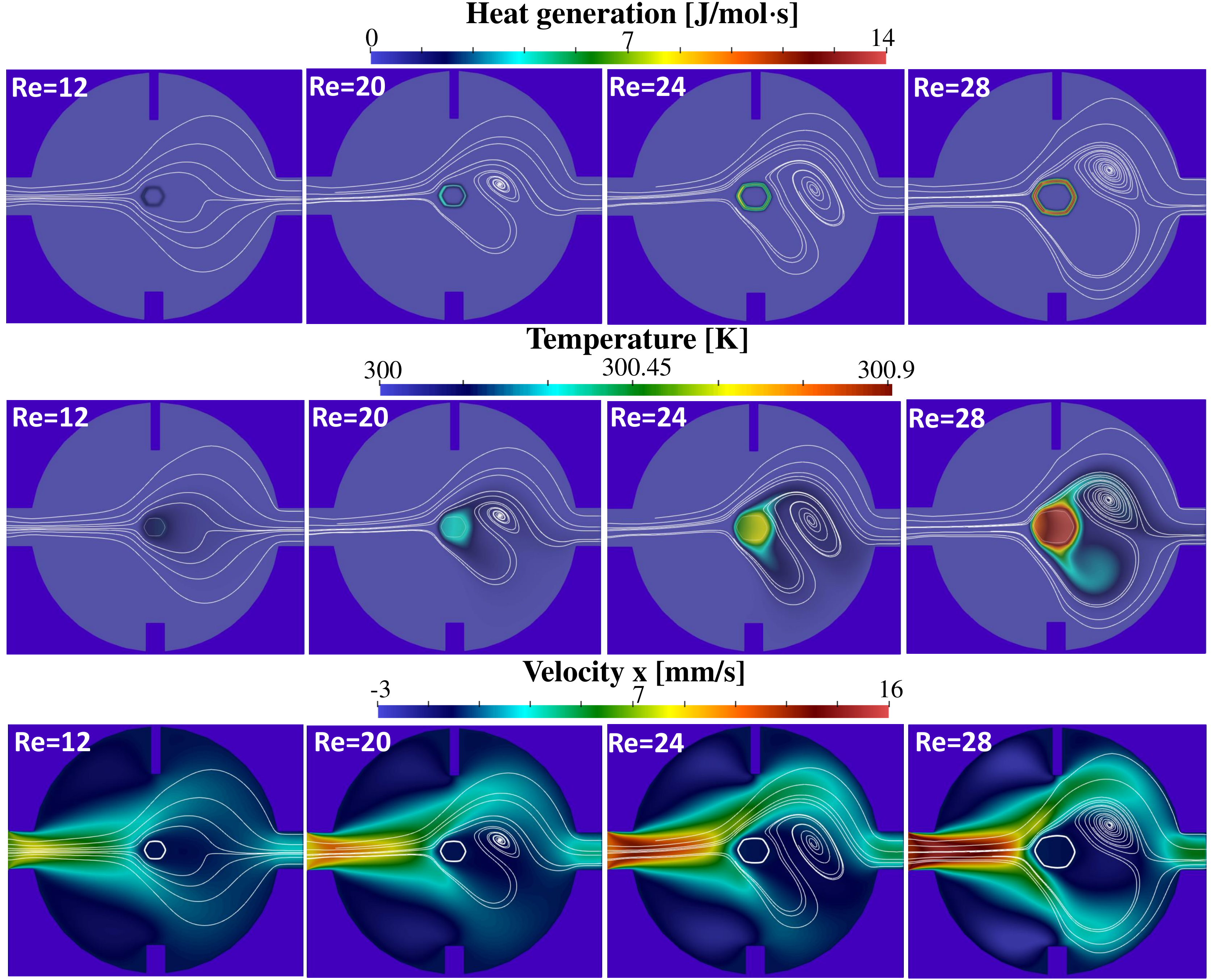}	
	\caption{Instantaneous fields of heat generation (top), temperature (center), velocity field (bottom) within the growth cell at time $t =$16 hours for different Reynolds numbers Re = 12, 20, 24, and 28 (from left to right), respectively.}
	\label{Reynold}
\end{figure}

\begin{figure}[!ht]                          
	\centering
	\includegraphics[width=0.6\textwidth]{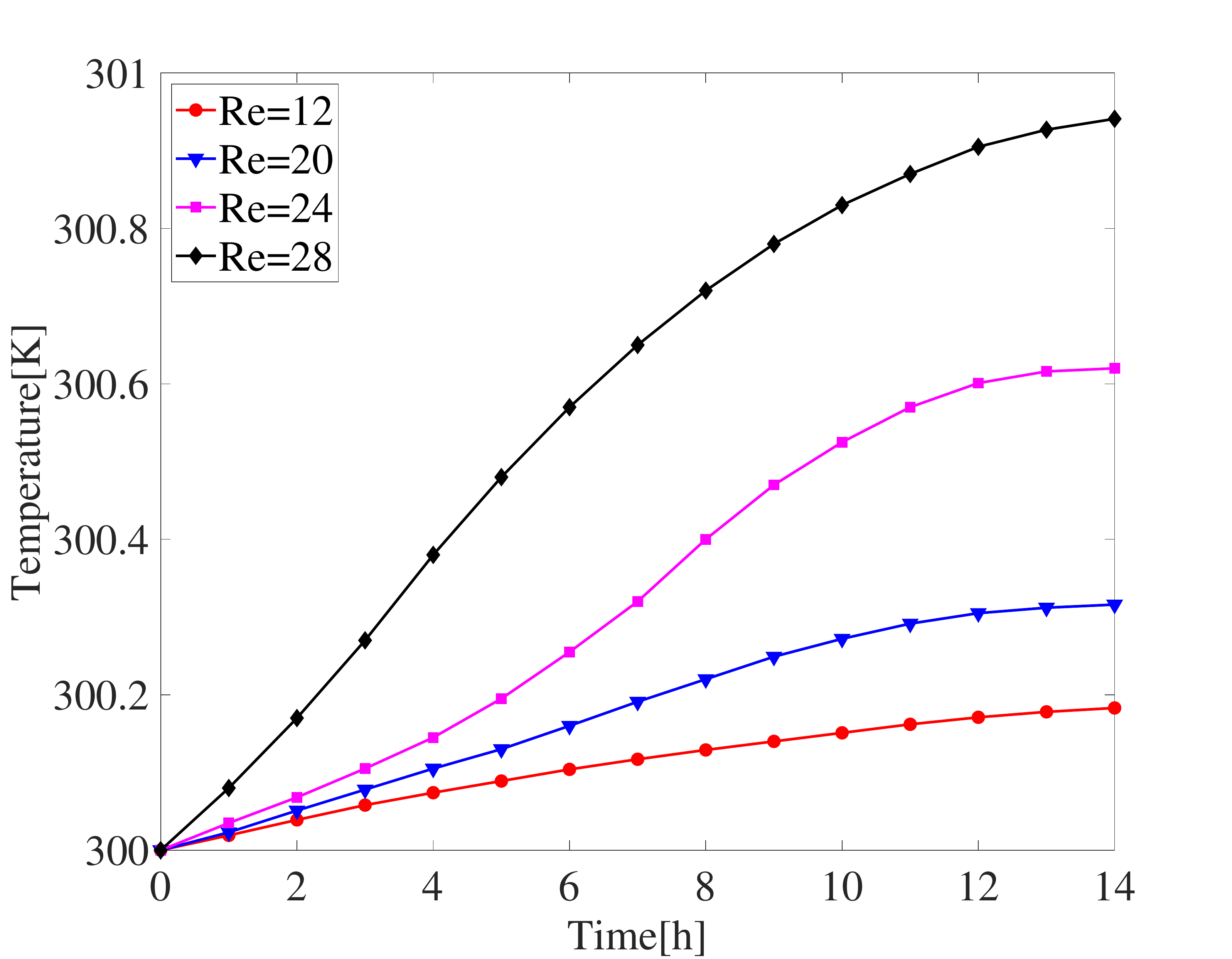}	
	\caption{Evolution of peak temperature with time within the cell for different Reynolds numbers Re = 12, 20, 24, and 28, respectively.}
	\label{Reynoldplot}
\end{figure}

In the real single-crystal reactor the incoming flow of (S)-mandelic acid in solution might have an impact on crystal growth rate and shape, as demonstrated in \cite{tan2022mandelic} when neglecting temperature changes. The aim of the present section is to check this point for different Reynolds numbers, and to suggest the inclusion of baffles to support symmetrical growth.\\
\paragraph{Effect of Reynolds number}
The Reynolds number is defined as Re$= \bm{u}_{in} D/\nu_f$, where $D$ is the initial diameter of the crystal seed, $\nu_f$ is the kinematic viscosity of water, taken at 1mm$^2$/s. The inlet velocity is set as $\bm{u}_{in}$ = 8, 10, 12 or 14 mm/s, respectively. \\
Figure~\ref{Reynold} shows that at higher Reynolds number, the crystal grows much faster. As a consequence, more heat is generated at the interface because of the intensive solute convection around the crystal. This effect dominates over the accelerated transport of heat away from the crystal by the flow. Overall, an increase of the maximum temperature with Re is observed around the single crystal (see Fig.~\ref{Reynoldplot}).

\begin{figure}[!ht]                          
	\centering
	\includegraphics[width=0.5\textwidth]{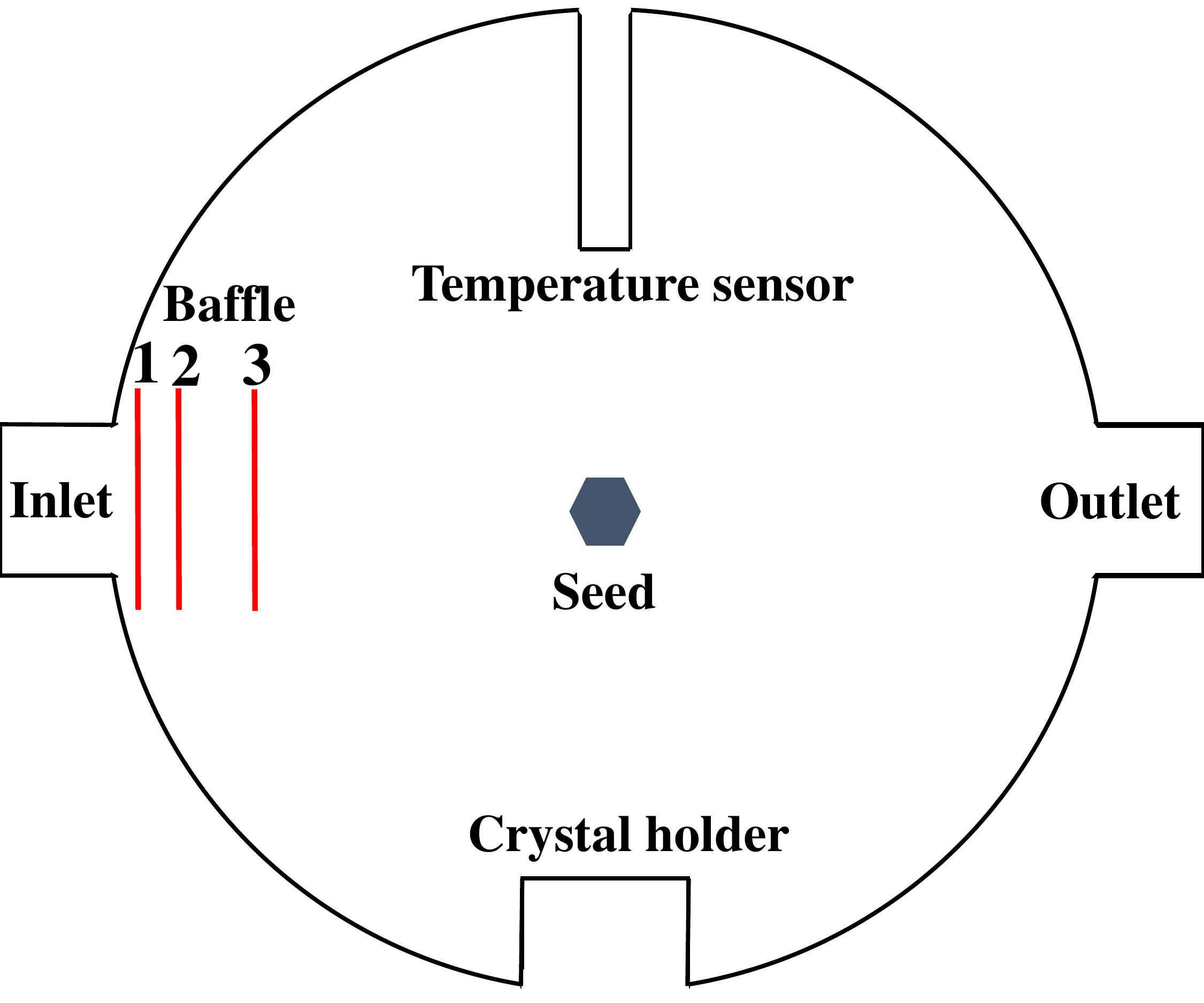}	
	\caption{2D growth-cell geometry including baffles at different positions.}
	\label{geobaffle}
\end{figure}

\paragraph{Baffle}
As seen from Fig.~\ref{Reynold} (white lines in the bottom figure, showing the crystal boundary), the overall shape of the crystal varies considerably as function of the Reynolds number, and rapidly becomes non-symmetric. However, the regularity of the crystal shape is a property of high interest regarding the performance of the final products. Therefore, it would be desirable to find a simple geometrical modification to the single-crystal growth cell, leading to isotropic growth rates and/or a desired final aspect ratio. For this purpose, a simple flat baffle has been included in the simulation domain in front of the inlet, in order to prevent a direct impact of the incoming flow onto the growing seed. Three different configurations (different positions) of the baffle have been compared. The resulting configurations are illustrated in Fig.~\ref{geobaffle}; what is called configuration 0 is the original case, without any baffle.\\

\begin{figure}[!ht]                          
	\centering
	\includegraphics[width=0.8\textwidth]{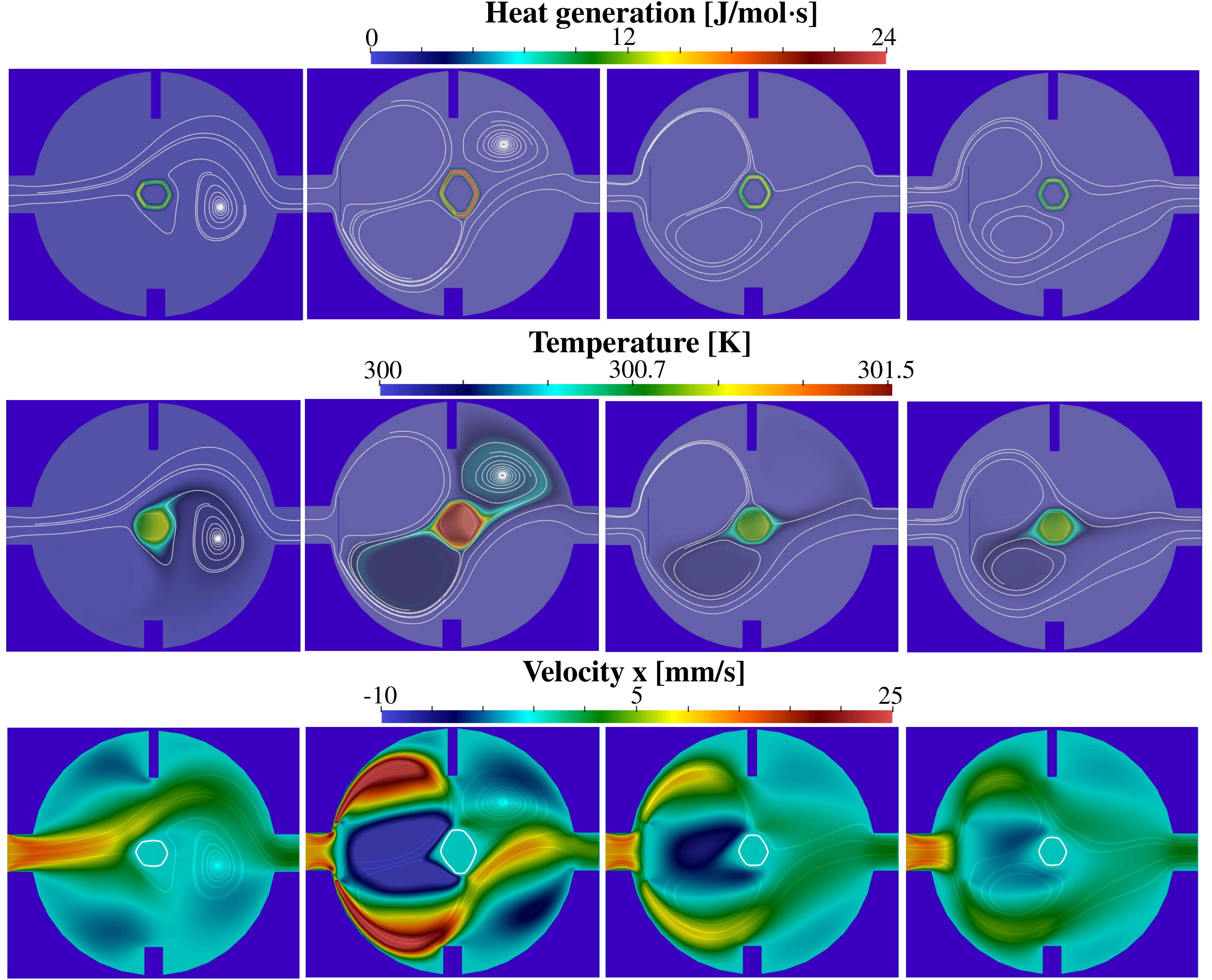}	
	\caption{Instantaneous fields of heat generation (top), temperature (center), velocity (bottom) at time $t =$ 16 hours in the growth cell with the baffle placed at different positions (from left to right): (1) without baffle; (2) with baffle at position 1; (3) with baffle at position 2; (4) with baffle at position 3.}
	\label{baffle_heat}
\end{figure}

Figure~\ref{baffle_heat} shows that ventilation effects are still visible with the baffle at position 1, much more than at other positions; this case leads to the faster crystal growth in vertical direction. The single crystal growth becomes more symmetric as the baffle is placed at a farther distance from the inlet of the growth cell. To quantify the effect of the baffles on the symmetry of the crystal, a quality parameter has been defined as $Q=\max(L_i)/\min(L_i)$ where index $i\in\{0,\dots,5\}$ covers the length of all sides of the resulting crystal. Thus, parameter $Q$ quantifies non-isotropic growth, with $Q=1$ (the minimum value) corresponding to a perfectly isotropic growth, while an increasing value of $Q$ corresponds to increasing non-isotropy. The values of crystal quality as obtained from all simulations after 16 hours of growth are listed in Table~\ref{baf}. Overall, the baffle in position 3 should be preferred to get maximum isotropy and minimum temperature effects.\\

\begin{table}[!htbp]
\centering
\setlength{\tabcolsep}{2.5mm}{
\caption{Impact of the different baffles (see Fig.~\ref{geobaffle}) on the isotropy ratio}\label{baf}
\begin{tabular}{|c|c|c|c|c|}
\hline
Position & No Baffle & Baffle 1 & Baffle 2 & Baffle 3\\
\hline
Q &   1.29 &  1.73 & 1.16  & 1.12 \\
\hline
\end{tabular}}
\end{table}

\begin{figure}[!ht]                          
	\centering
	\includegraphics[width=0.6\textwidth]{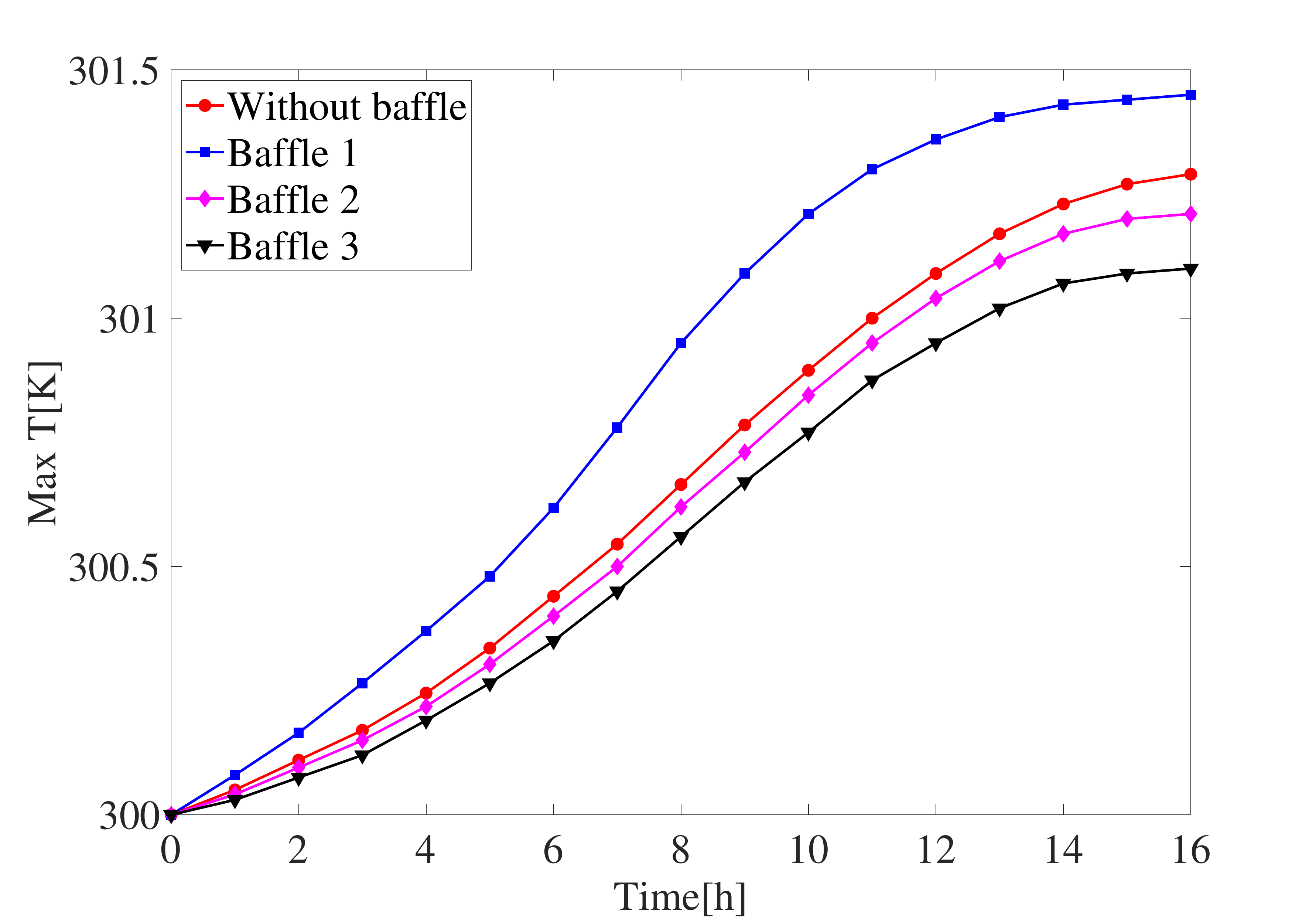}	
	\caption{Evolution of peak temperature with time for the baffles placed at different positions, for an initial temperature $T_0$ = 300K.}
	\label{baffle_velocity}
\end{figure}

Figure~\ref{baffle_velocity} shows the peak temperature as function of time for the different baffles. Baffle 1 corresponds to the large ventilation effects visible in Fig.~\ref{baffle_heat}; then, the crystal side facing the high flow velocity in vertical direction leads to a much larger growth rate there, generating much heat at the crystal interface.

\subsection{Conclusions and perspectives}
In this work, a hybrid \gls{lbm}/finite-difference method has been used to model the growth of a single crystal of (S)-mandelic acid. \gls{lbm} is used for the phase-field equation, while the finite-difference method is applied for the species and energy equations due to the high ratio between thermal and species diffusivity. Selected test-cases show that numerical stability can be achieved with the hybrid solver thanks to the finite-difference method. Successful verification and validation steps are documented. The results provide detailed information regarding the magnitude and dynamics of the temperature fields developing in the measuring cell during the growth process. The heat generation during phase change at the interface of the crystal leads overall to only small changes in temperature over the whole cell. These local changes in temperature lead to noticeable temperature gradients around the crystal. For all cases considered, a maximum temperature increase of almost 1.5$^{\circ}$C has been observed. In this particular case the molar heat generation at the interface can be probably neglected to address most questions of interest. However, convection can amplify temperature differences. Using a baffle located at a suitable position, ventilation and temperature effects can be minimized.

\section*{Acknowledgement}
The authors would like to acknowledge the financial support by the EU-program ERDF (European Regional Development Fund) within the Research Center for Dynamic Systems (CDS), as well as the computing time granted by the Universität Stuttgart-Höchstleistungsrechenzentrum Stuttgart (HLRS); all calculations for this publication were conducted with computing resources provided under project number 44216.

\end{document}